\documentclass[prb,aps,amssymb,amsmath,twocolumn,showpacs,floatfix]{revtex4-1}

\usepackage{graphicx}
\usepackage{bm}
\usepackage{amssymb,amsmath, mathrsfs}
\usepackage{color}
\usepackage{dsfont} 
\usepackage{amsfonts}

\begin{document}

\title{Quasiclassical theory of disordered multi-channel Majorana quantum wires}

\author{Patrick Neven, Dmitry Bagrets and Alexander Altland}

\affiliation{Institut f{\"u}r Theoretische Physik, Universit{\"a}t zu K{\"o}ln, K{\"o}ln, 50937, Germany}

\date{\today}
\pacs{74.78.Na, 71.23.-k, 73.63.Nm, 74.45.+c}

\begin{abstract}
  Multi-channel spin-orbit quantum wires, when subjected to a magnetic
  field and proximity coupled to $s$-wave superconductor, may support
  Majorana states.  We study what happens to these systems in the
  presence of disorder.  Inspired by the widely established
  theoretical methods of mesoscopic superconductivity, we develop \'a
  la Eilenberger a quasiclassical approach to topological nanowires
  valid in the limit of strong spin-orbit coupling.  We find that the
  ``Majorana number" ${\cal M}$, distinguishing between the state with
  Majorana fermion (symmetry class B) and no Majorana (class D), is
  given by the product of two Pfaffians of gapped quasiclassical
  Green's functions fixed by right and left terminals connected to the
  wire. A numerical solution of the Eilenberger equations reveals that
  the class D disordered quantum wires are prone to the formation of
  the zero-energy anomaly (class D impurity spectral peak) in the
  local density of states which shares the key features of the
  Majorana peak.  In this way we confirm the robustness of our
  previous conclusions [Phys.~Rev.~Lett. {\bf 109}, 227005 (2012)] on
  a more restrictive system setup. Generally speaking, we find that
  the quasiclassical approach provides a highly efficient means to
  address disordered class $\mathrm{D}$ superconductors both in the
  presence and absence of topological structures.
\end{abstract}

\maketitle

\section{Introduction}

Semiconductor quantum wires proximity coupled to a conventional
superconductor and subject to a magnetic field may support Majorana
fermion edge states~\cite{Lutchyn:2010,Oreg:2010}. Building on
relatively conventional device technology, this proposed realization of
the otherwise evasive Majorana fermion has triggered a wave of
theoretical and experimental activity, which culminated in the recent report
of a successful observation by several experimental
groups~\cite{Mourik:2012, Deng:2012, Heiblum:2012}. In these
experiments, evidence for the presence of a Majorana is drawn from the
observation of a zero bias peak in the tunneling conductance into the
wire. While the observed signal appears to be naturally explained in
terms of a Majorana end state, two of us have pointed out that a
midgap peak might be generated by an unrelated
mechanism~\cite{Bagrets:2012}: in the presence of even very moderate
amounts of disorder, the semiconductor wire supports a zero energy ``spectral peak"
(an accumulation of spectral weight at zero energy) which
resembles the Majorana peak in practically all relevant
aspects. Specifically, it
is (i) rigidly locked to zero energy,
 (ii) is of narrow width of $\mathcal{O}(\delta)$, where $\delta$ is
  the single particle level spacing, (iii)  
carries integrated spectral weight $\mathcal{O}(1)$, and
(iv) relies on parametric conditions (with regard to spin orbit
   interaction, proximity coupling, magnetic field, and chemical potential) identical to
   those required by Majorana state formation.
What makes the spectral peak \textit{distinct} from the Majorana is that it
relies on the presence of a moderate amount of disorder,
viz. impurity scattering strong enough to couple neighboring single particle
Andreev levels. Besides, the spectral peak is
vulnerable to temperature induced dephasing. While this marks a
difference to the robust Majorana state, the reported
experimental data \textit{does} show strong sensitivity to temperature, which may
either be due to an intrinsic sensitivity of the peak, or due to a
temperature induced diminishing of the measurement sensitivity, or
both. In either case, the situation looks inconclusive in this
regard. 

Generally speaking, the results of Ref.~\onlinecite{Bagrets:2012}, as
well as those of Refs.~\onlinecite{Potter:2012,Beenakker:2012} suggest
that the observation of a midgap anomaly in the tunneling conductance
might be due to either mechanism, disorder peak, Majorana peak, or a
superposition of the two, and this calls for further research.

Our previous study was based on an analytically tractable idealization
of a semiconductor quantum wire subject to a magnetic field sweep. In
the present paper, we will explore the role of disorder scattering
within a model much closer to the experimental setup. The price to be
payed for this more realistic description is that a fully analytic
treatment is out of the question. Instead, we will employ a
semi-analytic approach based on the formalism of quasiclassical Green
functions. Introduced in the late
sixties~\cite{Eilenberger:1968,Larkin:1969}, the latter has become an
indispensable tool in mesoscopic superconductivity~\cite{Rammer:1986,
  Eschrig:2000, Altland:2000, Feigelman:2000} and quantum transport in
general~\cite{Nazarov:book}. We here argue that quasiclassical methods
are, in fact, tailor made to the modeling of Majorana quantum wires
(or, more generally, quasi one-dimensional topological
superconductors.) Specifically, we will show that in the problem at
hand, the quasiclassical ``approximation'' is actually very
mild. Further, the quasiclassical Green's function affords a
convenient description of the topological signatures of the system in
terms of Pfaffians. Finally, the approach can be applied to systems
for a given realization of the disorder, and at numerical cost much
lower than that of exact diagonalization approaches. As a result, we
will be in a position to accurately describe local spectral properties
within a reasonably realistic model of a topological multi-channel
superconductor. As we are going to discuss below, our findings support
the principal statement made in Ref.~\onlinecite{Bagrets:2012}.

The rest of the paper is organized as follows. In
section~\ref{sec:qualt-disc-results}, we discuss the principal role
played by disorder in the system, the idea of the quasiclassical
approach, and its main results. In section~\ref{model} we specify our
model system, the quasi-classical approach is introduced in
section~\ref{quasi}, and in \ref{num} we discuss the numerical
solution of the quasiclassical equations. We conclude in
section~\ref{concl}.  A number of technical details are relegated to
Appendices.

\section{Qualtiative discussion and results}
\label{sec:qualt-disc-results}

\begin{figure}[t]
 \includegraphics[width=8cm]{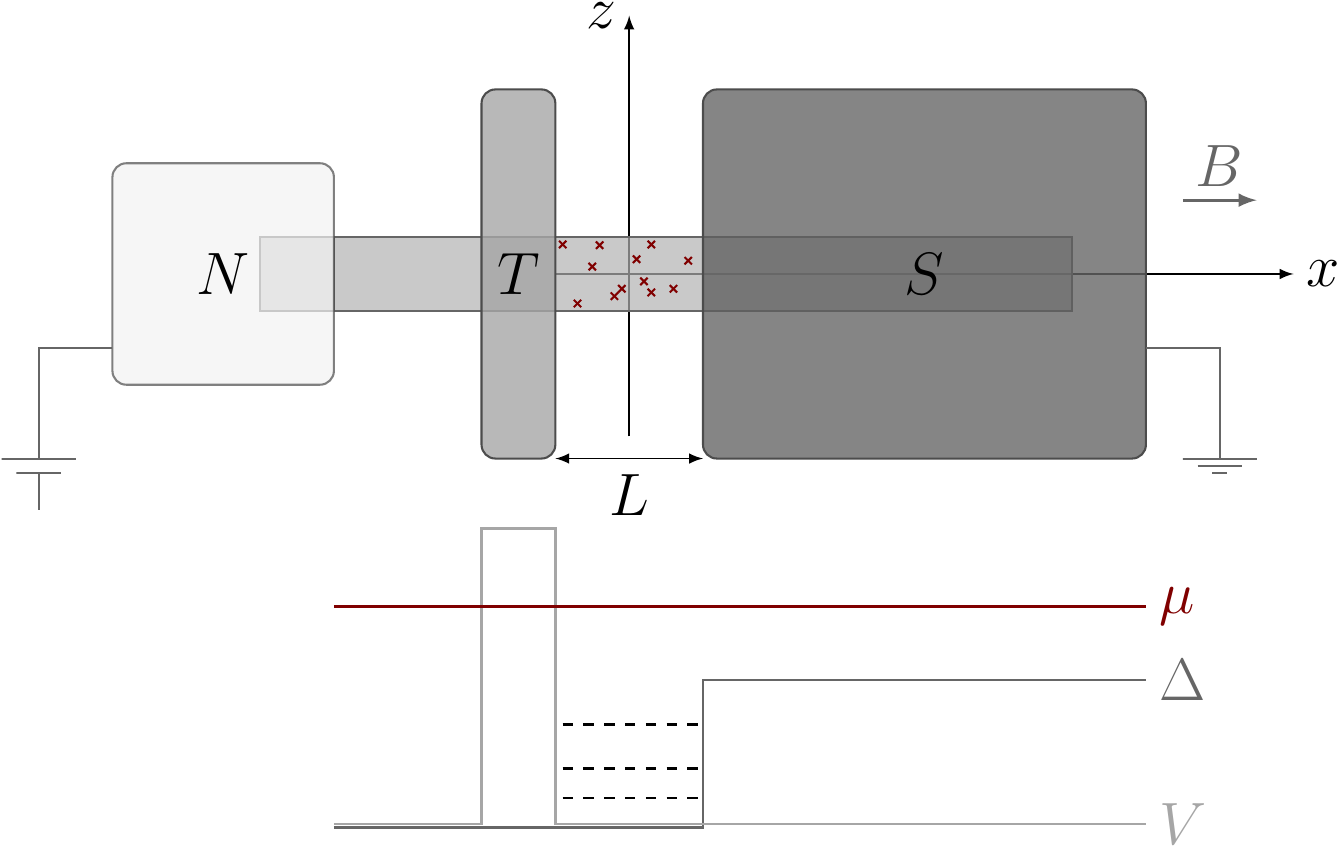}
 \caption{Disordered spin-orbit nanowire, subjected to colinear
   magnetic field, is proximity coupled to the $s$-wave superconductor
   (S) and terminated by the tunneling barrier (T) at one of its ends.
   A sketch below shows the profile of the induced superconducting gap
   $\Delta(x)$ and gate induced potential $V(x)$ defining the tunnel
   barrier. Andreev bound states (ABS) are depicted by dotted lines.}
\label{fig:Setup}
\end{figure}

A schematic of the device currently under experimental investigation
is shown in Fig. \ref{fig:Setup}. A semiconductor quantum wire
subjected to strong spin orbit interaction is brought in contact to a
superconductor (S), and, via a tunnel barrier (T) to a normal metal
lead (N). The application of a small excess voltage, $V$, to the
latter induces a tunnel current into the central region. The
differential conductance $dI/dV$ probes the (tunneling) density of
states at an energy $V$ (units $e=\hbar=c=1$ throughout) relative to
the systems chemical potential, $\mu$. The physics we are interested
in is contained in a band center $(V=0)$ anomaly in that quantity.

In a manner to be discussed in more detail below, one contribution to
the band center density of states is provided by a Majorana bound
state localized at the tunnel barrier. The second, spectral
contribution is generated by a conspiracy all other low lying
quasiparticle states in the system. Technically, these are Andreev
bound states forming at energies $\pm E_j$ in the region between the
tunnel barrier and the superconductor. The number of these states
increases with the extension of the wire -- a few hundred nanometers
in the experiment -- and the number of transverse channels below the
chemical potential. In the presence of disorder, the entity of these
states defines an effective ``quantum dot". The proximity of a
superconductor, and the breaking of both spin rotation and time
reversal symmetry imply that the system belongs to symmetry class
D~\cite{Altland:1997}.

The symmetry of class D random systems implies a clustering of levels
at zero energy. Loosely speaking, the conventional level repulsion of
random spectra turns into a zero energy level attraction. On a
resolution limited to scales of order of the mean level spacing, the
zero energy density of states (DoS) is enhanced by a factor of two relative to the mean
background, i.e. it shows a peak. This phenomenon manifests itself
at relatively small sample-to-sample fluctuations, i.e. the peak is a
sample specific effect. In passing we note that the weak
anti-localization phenomenon discussed in Ref.~\onlinecite{Beenakker:2012}
rests on the same principal mechanism of midgap quantum interference.

Below, we will explore the phenomenon of spectral peak formation in a
setting modelled to closely mimic the ``experimental reality". To be
more specific, we consider a semiconductor wire supporting a number of
$N>1$ transverse channels below the chemical potential. We assume a
value of the chemical potential such that the highest lying of these
channels is ``topological" (chemical potential falling into the gap
opened by the simultaneous presence of order parameter and magnetic
field.)

To quantitatively describe this phenomenon, we generalize the
quasiclassical Green function approach to the present setting. Indeed,
the phenomena we are interested in manifest themselves on length
scales large in comparison to the Fermi wavelength, yet smaller than
the relevant coherence length, and this makes them suitable to
quasiclassical treatment.  The quasiclassical theory of the present
paper is formulated in terms of the Eilenberger function $\tilde
Q(x;\epsilon)$, which is a position and energy dependent matrix of
size $8N\times 8N$. The factor of $8=2\times 2\times 2$ accounts for
spin, chiral (left/right modes) and particle-hole degrees of freedom.
It will turn out that the structure of the theory is most
transparently exposed in the so-called Majorana basis, where the
Eilenberger function becomes a real antisymmetric matrix. The Pfaffian
of that matrix, ${\rm Pf}(\tilde Q)$ will be seen to assume two values
($\pm 1$), which locally (in space) identify the $\mathds{Z}_2$
invariant of the underlying one-dimensional class $\mathrm{D}$
superconductor, in dependence on system parameters: for values of the
magnetic field smaller or larger than a critical field, $B_c\equiv
\sqrt{\Delta^2+\mu^2}$, where $\mu$ and $\Delta$ are chemical
potential and bulk order parameter, the invariant is trivial or
non-trivial, resp. In the latter case, the wire supports a Majorana
state at the barrier, in the former it doesn't.

\begin{figure}[t]
\includegraphics[width=7cm]{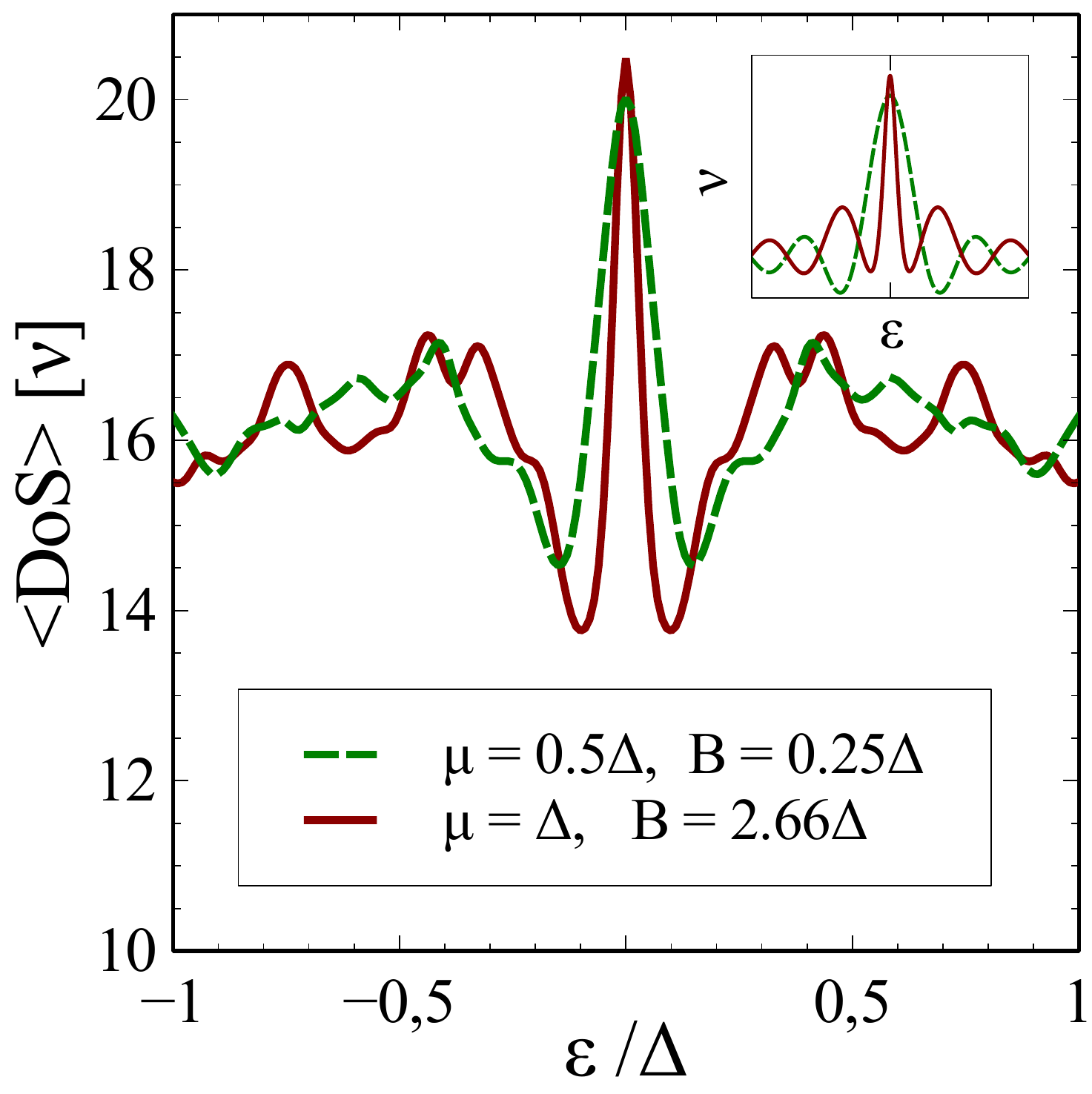}
\caption{Disorder averaged local density of states (LDoS) at the left
  end of the spin-orbit quantum wire sketched in Fig.~\ref{fig:Setup}
  (in units $\nu=1/2\pi v$).  The number of occupied bands $N=2$
  corresponds to four transport channels.  Parameters are: (i) red
  solid line (${\cal M} = -1$), $B=2.66 \Delta$, $\mu= \Delta$ ---
  topological phase; (ii) green dashed line (${\cal M} = +1$), $B=0.25
  \Delta$, $\mu= 0.5 \Delta$ --- trivial phase. The wire length $L = 4
  v/\Delta$, dimensionless strength of disorder $\gamma_w^2/v = 0.16
  \Delta$, which translates into the mean free path $l=0.4 L$.
  Tunneling rate $\Gamma = 5\cdot 10^{-2} \Delta$.  Velocities in two
  bands were taken to be equal, $v_1=v_2=v$. The inset shows profiles
  of the DoS resulting from random matrix theory.}
\label{fig:LDOS_average}
\end{figure}

Fig.~\ref{fig:LDOS_average} shows a profile of the local DoS (LDoS)
computed at the left end of the wire for typical system parameters as
detailed below. The green dashed and red solid curve correspond to a
situation without and with Majorana state. Within our model, the
states in the wire carry a narrow width, $\sim g_T \delta$, reflecting
the possibility of decay through the tunnel barrier into the adjacent
lead. (Here, $\delta$ is the mean level spacing of the wire, and
$g_T\lesssim 1$ the barrier tunneling conductance.) This broadening
accounts for the finite width of the Majorana peak in the topological
regime. Loosely speaking, the negative shoulders observable next to
the center peak are due to the repulsion of adjacent levels of the
center Majorana level. A more substantial explanation is as follows:
for an odd parity of the total number of levels --- a signature of the
topological regimes --- the disordered quantum system falls into
symmetry class $\mathrm{B}$, rather than $\mathrm{D}$. (Class
$\mathrm{B}$ is the designation for a system with the same symmetries
as a class $\mathrm{D}$ system, yet odd number of levels.) A universal
signature of class $\mathrm{B}$ is a \textit{negative} spectral peak
at zero energy (the negative of the positive class $\mathrm{D}$ peak),
superimposed on a single $\delta$-peak (the Majorana). The joint
signature of these two structures is seen in the solid curve in
Fig. \ref{fig:LDOS_average}.

At resolutions limited to values $\sim \delta$, the superposition of
the Majorana and the class $\mathrm{B}$ peak looks next to
indistinguishable from the class $\mathrm{D}$ peak (dashed), and this
similarity of unrelated structures might interfere with the
unambiguous observation of the Majorana by tunneling spectroscopy.
Indeed, the differential tunneling conductance monitored in
experiment,
\begin{equation}
\frac{dI}{dV} = \frac{e^2\,  g_T }{16\pi\hbar} \int_{-\infty}^{\infty} 
\frac{\partial f_F}{\partial \epsilon}(\epsilon-V)\,\frac{\nu_L(\epsilon)}{\nu} \mathrm{d}\epsilon,
\label{eq:dI_dV}
\end{equation}  
is essentially~\footnote{A refined variant of
  Eq.~(\ref{eq:dI_dV}) has recently been derived in
  Ref.~\onlinecite{Ioselevich:2012}. It was found that for extremely
  low temperatures, $T\ll \Gamma$ the tunneling current may contain a
  dip reflecting the mutual cancellation of electron and hole current
  contributions. Our discussion is thus tacitly assumes $T\gtrsim
  \Gamma$.} determined by the local density of states, i.e. the
structures shown in Fig. \ref{fig:LDOS_average} are expected to
reflect directly in the measured signal. (Here, $f_F$ is the Fermi
distribution, $\nu_L$ is the DoS at the left barrier, $\nu$ is the DoS in the single chiral 
channel per unit length, and $V$ the applied voltage.)

The profiles of the curves shown in Fig. \ref{fig:LDOS_average} were
computed for a two-channel quantum wire at a mean free path $l\simeq
L$ of the order of the system size. We are addressing a system at the
interface between the ballistic and the localized regime. In view of
these system parameters it is remarkable that the DoS profiles in
Fig. \ref{fig:LDOS_average} show striking similarity to the average
DoS of a class \textrm{D} and \textrm{B} \textit{random matrix}
model~\cite{Altland:1997}. For comparison the average DoS of a class
$\mathrm{B}$ and $\mathrm{D}$ random matrix Hamiltonian is shown as an
inset in Fig. \ref{fig:LDOS_average}. The similarity of the results
indicates that the system of subgap states in our system behaves as if
it formed an effective chaotic quantum dot localized in the vicinity
of the left system boundary (right to the tunnel barrier).  Taking
into account spin, chiral and channel quantum numbers, the mean level
spacing in such dot is given by $\delta \simeq \pi v/2NL$. In the
presence of magnetic field the BCS gap in the superconducting region
of the wire reads $\epsilon_- = |B-\sqrt{\Delta^2 + \mu^2}|$.  The
number of subgap Andreev levels forming the effective dot is thus
given by $N_{\rm levels} \simeq 2\epsilon_-/\delta$.

\begin{figure}[t]
\includegraphics[width=7cm]{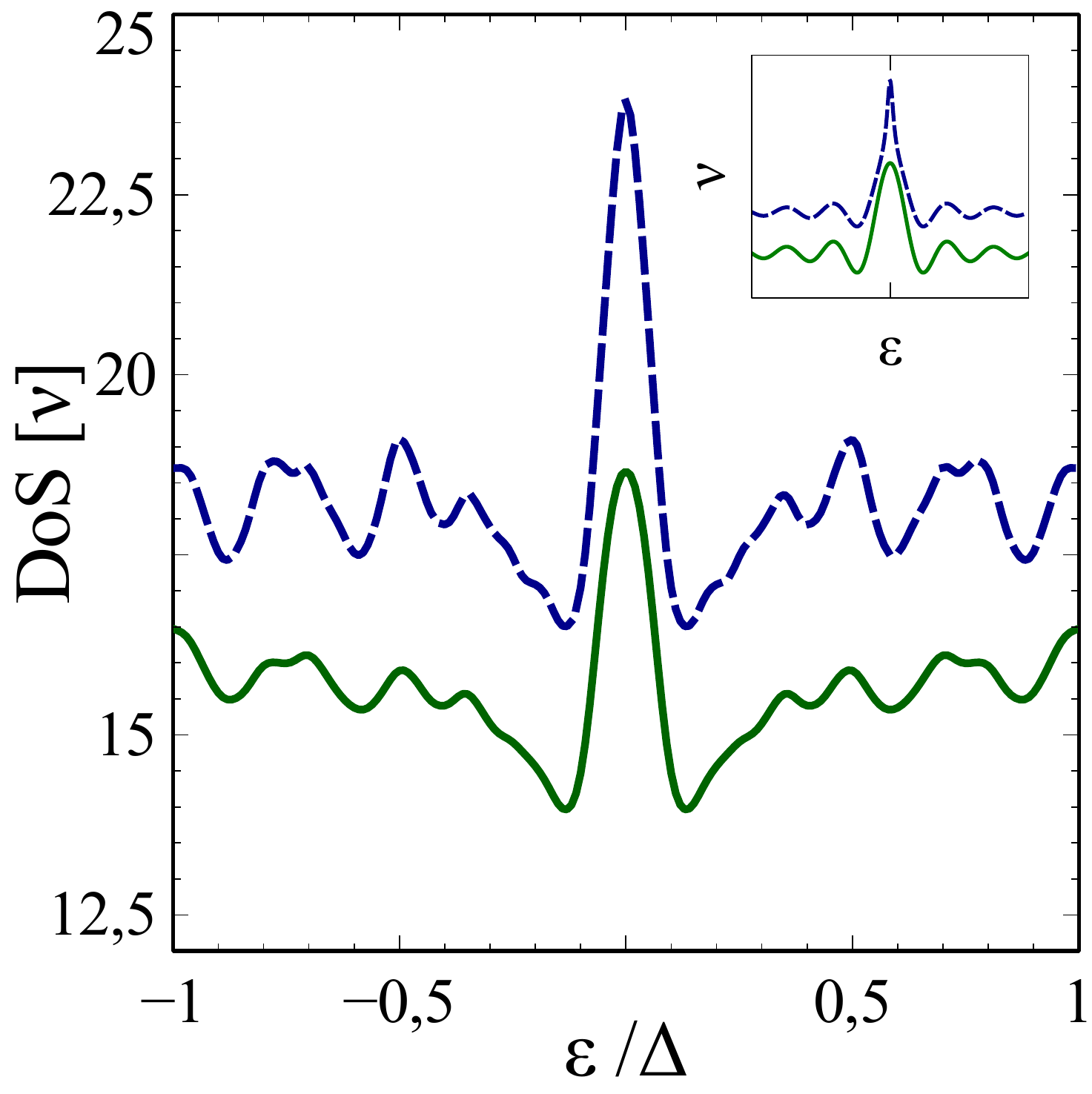}
\caption{The average LDoS (solid green line) and the square root of
  its second (reducible) moment $\langle
  \nu^2_L(\epsilon)\rangle^{1/2}$ (dashed blue line) at the left end
  of the spin-orbit quantum wire in the trivial phase (${\cal M}
  =+1$).  System parameters are listed in
  Fig.~\ref{fig:LDOS_average}.  The inset shows profiles of the mean 
  DoS and the square root of the two level correlation function resulting 
  from random matrix theory.
}
\label{fig:LDOS_variance}
\end{figure}

The profiles shown in Fig. \ref{fig:LDOS_average} are ensemble averages,
$\langle \nu_L\rangle$ of the LDoS, $\nu_L$, 
where the sampling was over $\sim 500$ randomly chosen impurity
configurations. To demonstrate the weakness of fluctuations,
Fig. \ref{fig:LDOS_variance} compares the average LDoS (solid line) to
the ``typical'' LDoS, i.e. the average $\sqrt\langle \nu_L^2\rangle$. The
relatively minor deviation between average and typical DoS
demonstrates that the standard deviation
\begin{equation}
\delta \nu(\epsilon) = \Bigl\langle \Bigl(\nu(\epsilon)- 
\langle\nu(\epsilon)\rangle\Bigr)^2\Bigr\rangle^{1/2}
\end{equation}
characterizing the strength of mesoscopic fluctuations is relatively
small.

\begin{figure}[t]
\includegraphics[width=7cm]{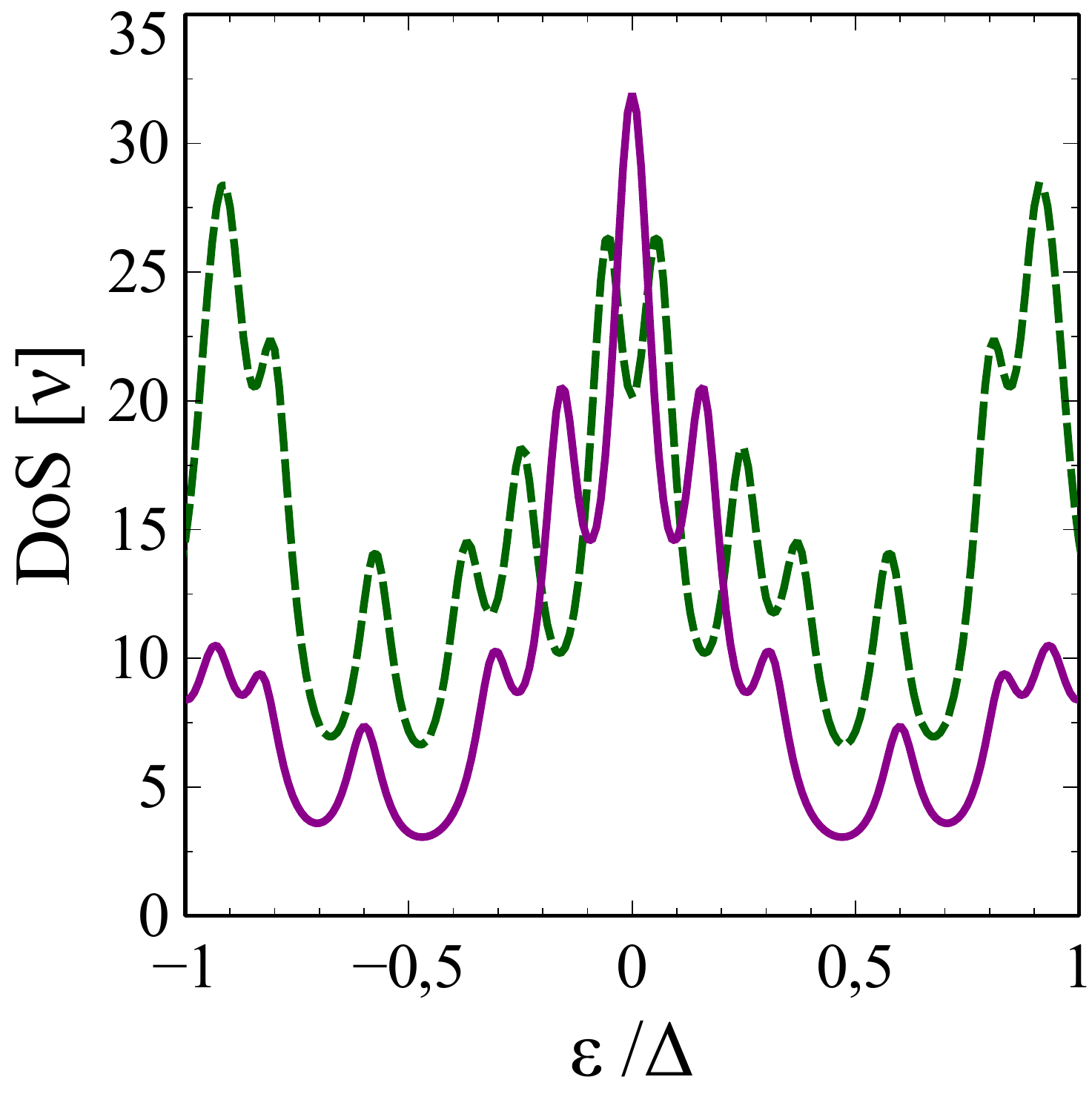}
\caption{The sample specific LDoS in the trivial phase without
  Majorana state (${\cal M}=+1$) for two different disorder
  realizations. Tunneling rate $\Gamma = 0.05\Delta$, other system
  parameters are the same as in Fig.~\ref{fig:LDOS_average}.  Curves
  demonstrate two typical scenarios: (i) two conjugate Andreev bound
  states $\pm \epsilon_{\rm min}$ lying close to Fermi energy and
  having energy splitting $\sim \Gamma$ (dotted green line); (ii)
  particle and hole states have merged into a single zero-energy peak
  of width $\Gamma$ and can not be resolved by tunnel spectroscopy
  (solid magenta line). For the chosen set of parameters the mean
  level spacing $\delta = 0.2 \Delta$ and the gap in the S region
  $\epsilon_- = 0.87 \Delta$. Thus one has approximately $N_{\rm levels} \simeq 8$
  random Andreev levels.}
\label{fig:LDOS_sample}
\end{figure}

The weakness of fluctuations implies that the disorder peak
is a realization specific phenomenon. This is demonstrated explicitly
in Fig.~\ref{fig:LDOS_sample}, where two un-averaged DoS profiles
individual disorder configurations are shown. The data is for a
``non-topological" system, no non-degenerate zero energy level
is present. Depending on whether the pair of lowest lying Andreev states
$(+\epsilon_\mathrm{min},-\epsilon_\mathrm{min})$ exceeds the level
broadening $\Gamma$, the DoS enhancement will assume the form of a
single peak (magenta solid), or a split peak (green dashed). In either
case, an excess DoS of integrated spectral weight $\sim 1$ is
present.  

We finally note that both the Majorana peak and the $\mathrm{D}$ spectral peak crucially rely
on the presence
of a magnetic field. In the absence of a field
time-reversal symmetry is restored, and the wire turns into a member
of symmetry class $\mathrm{DIII}$. Such systems display a DoS
depletion at zero energy, rather than a peak. In other words, the  spectral peak discussed here 
will disappear along with  the Majorana resonance.

In the rest of the paper we discuss how the results summarized here
were obtained by a combination of quasiclassical and numerical
methods. 
 
\section{Model of the Majorana nanowire}
\label{model}

We consider a multi-band quantum wire of width $L_z$,
subject to Rashba spin-orbit 
coupling, proximity coupling to 
an $s$-wave superconductor, and to a magnetic
field~\cite{Lutchyn:2011}. We choose coordinates such that the wire lies
along the $x$-axis, parallel to the magnetic field, the $y$-axis is
perpendicular to the surface of the superconductor, and 
the spin orbit field is pointing along the $z$-axis.

In the rest of this section we will specify the Bogoliubov-de-Gennes
(BdG) Hamiltonian describing this system in the presence of disorder
(section~\ref{sec:BdG}). We will then linearize the electron spectrum
in the limit of strong spin-orbit coupling thereby introducing a
description in terms of right and left one-dimensional chiral fermions
(section~\ref{sec:one-dim-chiral-ferm}), and finally transform the
Hamiltonian into Majorana basis
(section~\ref{sec:Majorana-rep}). Experts in the description of
topological quantum wires may proceed directly to the end of the
section.

\subsection{Bogoliubov-de-Gennes Hamiltonian}
\label{sec:BdG}
Introducing a four component spinor $\Psi = (\psi_\uparrow,\psi_\downarrow, \bar\psi_\uparrow,\bar\psi_\downarrow)$
in the product of spin and particle-hole spaces, the
BdG Hamiltonian $\cal H$ describing the system in the $xz$-plane reads
\begin{equation}
{\cal H} = \frac{1}{2}\int \bar\Psi(x,z)\left(
\begin{array}{cc}
\hat h_0  + \hat W & i \hat s_y \Delta^* \\
-i \hat s_y \Delta & -\hat h_0^\mathrm{T} - \hat W^\mathrm{T}
\end{array}\right)\Psi(x,z) \mathrm{d}x\, \mathrm{d}z,
\label{eq:H_full_xz}
\end{equation}
with 
\begin{equation}
\hat h_0 = -(\partial_x^2+\partial_z^2)/2m - \mu(x)  + B(x) \hat s_x - i \alpha ( \hat s_z \partial_x  - \hat s_x\partial_z ).
\nonumber
\end{equation}
Here, $\Delta=\Delta(x)$ is the proximity amplitude induced the
superconductor, $B\equiv \frac 12 g \mu_B H$, where $H$ is the
external field, we have taken into account the transverse momentum
($-i\partial_z$) in the Rashba term, and $\hat W(x)$ is the random
disorder Hamiltonian. The Pauli matrices $\hat s$ operate in spin
space.

We proceed by introducing a system of transverse wave functions,
$\{\Phi^\sigma_n(z)\}$, which leads to a linear decomposition of the
Grassmann fields ($\sigma=\uparrow,\downarrow$) as
\begin{equation}
 \psi_\sigma = \sum_{n\sigma} \Phi_n^\sigma(z) \psi_\sigma^{(n)} (x), \quad 
 \bar\psi_\sigma = \sum_{n\sigma} \Phi_n^\sigma(z)^* \bar\psi_\sigma^{(n)} (x).
 \end{equation}
Defining
\begin{eqnarray}
\Psi^{(n)}&=&(\psi^{(n)}_{\uparrow}, \psi^{(n)}_{\downarrow}, \bar\psi^{(n)}_{\uparrow}, \bar\psi^{(n)}_{\downarrow})^\mathrm{T}, \nonumber \\
\bar\Psi^{(n)}&=& (\bar\psi^{(n)}_{\uparrow}, \bar\psi^{(n)}_{\downarrow}, \psi^{(n)}_{\uparrow}, \psi^{(n)}_{\downarrow}), 
\end{eqnarray}
the Hamiltonian then takes the form
\begin{widetext}
\begin{equation}
{\cal H} = \frac{1}{2}\int \bar\Psi^{(n)}(x)\left(
\begin{array}{cc}
\hat h^{(n)}_0 \delta_{nm}  + (i\alpha \hat s_x\partial_z)_{nm} + \hat W_{nm} & i \hat s_y \Delta^*\delta_{nm} \\
-i \hat s_y \Delta\delta_{nm} & -(\hat h_0^{(n)} )^\mathrm{T}\delta_{nm} - (i\alpha \hat s_x\partial_z)_{mn} - \hat W^\mathrm{T}_{mn}
\end{array}\right)\Psi^{(m)}(x) \mathrm{d}x,
\label{eq:H_full_x}
\end{equation}
\end{widetext}
where the one-dimensional Hamitonian $\hat h^{(n)}_0$ acts in the
$n$-th band as
\begin{equation}
\hat h^{(n)}_0 = -\partial_x^2/2m +\mu_z^{(n)}  - \mu(x)  + B(x) \hat s_x - i \alpha \hat s_z \partial_x.
\end{equation}
In the case of an ideal waveguide the transverse wavefunctions are
$\Phi_n^\sigma(z) = \sqrt{2/L_z} \sin( n \pi z/L_z) $, so that
$\mu_z^{(n)} = \pi^2 n^2/ (2m L_z^2)$ and the matrix elements of the
spin-orbit interaction read
\begin{equation}
h_{nm}^{\rm s.o.} = (i\alpha \hat s_x\partial_z)_{nm}  = -\frac{2i\alpha}{L_z}\frac{nm}{n^2-m^2}\,(1-(-1)^{n+m}) \, \hat s_x.
\end{equation}
Let us now assume a thin wire, $L_z\lesssim l_{\rm so} =
\hbar/(m\alpha)$, where $ l_{\rm so} $ is the spin-orbit length.  In
this case the matrix elements $h_{nm}^{\rm s.o.} \ll \mu_z^{(n)}$ can
be treated as perturbations.  To this end we introduce a unitary
transformation $\hat {\cal U}$, which brings the high energy part of
the Hamiltonian, $(\mu_z^{(n)} \delta_{nm} + h_{nm}^{(n)})$, to
diagonal form. Due to the weakness of the perturbation, the
transformation is close to unity, $\hat {\cal U}=\exp( i \hat
X) \simeq 1 + i \hat X$, with generators $\hat X_{nm} = {\cal
  O}(L_z/l_{so})$. To first-order perturbation theory their
explicit form reads
\begin{equation}
\hat X_{n\neq m} \simeq \frac{i h^{\rm s.o.}_{nm}}{ \mu_z^{(m)} - \mu_z^{(n)}}, \qquad \hat X_{nn} = 0.
\end{equation}
The transformation $\Psi \to \hat {\cal U} \Psi$ and
$\bar\Psi \to \bar \Psi \hat {\cal U}^\dagger$  generates the approximate
Hamiltonian
\begin{widetext}
\begin{equation}
{\cal H} \simeq \frac{1}{2}\int \bar\Psi^{(n)}(x)\left(
\begin{array}{cc}
\hat h^{(n)}_0 \delta_{nm} +  \hat V_{nm} + \hat W_{nm} & i \hat s_y \Delta\delta_{nm} + (\delta\hat\Delta)_{nm} \\
-i \hat s_y \Delta\delta_{nm} - (\delta\hat\Delta)^*_{nm} & -(\hat h_0^{(n)} )^\mathrm{T}\delta_{nm} -  \hat V^\mathrm{T}_{mn} - \hat W^\mathrm{T}_{mn}
\end{array}\right)\Psi^{(m)}(x)\,\mathrm{d}x.
\label{eq:H_approx_x}
\end{equation}
\end{widetext}
Here we have redefined our notation for the disorder potential $\hat W
\to \hat{\cal U}\hat W \hat{\cal U}^\dagger$, and introduced a small
correction to the quasiparticle Hamiltonian, $\hat V = \alpha\, [\hat
X, \hat s_z]\, \partial_x $, and to the order parameter,
$\delta\hat\Delta = - \Delta [ \hat X, \hat s_y ]$.  One can estimate
the matrix elements of these operators as $V_{nm} \sim \epsilon
(L_z/l_{\rm so})$ and $(\delta\Delta)_{nm} \sim \Delta (L_z/l_{\rm
  so})$, where $\epsilon$ is the quasiparticle energy.  Since we are
primary interested in the effects of disorder, we will limit our
discussion to the situation when the low-energy physics is disorder
dominated, i.e.  the random potential $\hat W > {\rm max}\{\hat V,
\delta \hat \Delta\}$ masks the off-diagonal matrix elements of the
deterministic Hamiltonian. Having this in mind we thus omit both $\hat
V$ and $\delta \hat \Delta$ throughout the paper. However, this
assumption does not affect our results in qualitative ways.

Assume that the chemical potential lies close to the bottom of the
$N$-th band, $\mu \simeq \mu_z^{(N)}$.  We refer to this band as the
``topological band", since in the absence of interband scattering it
defines whether the quantum wire is in the topological phase or not.
The condition of the topologically non-trivial phase reads $B^2> B^2_c
= \Delta^2 + (\mu-\mu_z^{(N)})^2$.  We also assume a hierarchy of
energy scales $\mu_z^{(N)} - \mu_z^{(N-1)} \gg E_{\rm so} \gg \Delta
\sim B$, where $E_{\rm so} = m\alpha^2/2$ is the spin-orbit
energy. This condition implies that all other channels with the band
index $n<N$ are in the trivial phase, since for them $B^2 \ll \Delta^2
+ (\mu-\mu_z^{(n)})^2$. In the current experiments $E_{\rm so}\gtrsim
\Delta$, i.e. our theory is on the border of applicability (see,
however, the discussion in section~\ref{quasi}).

\subsection{One-dimensional chiral fermions}
\label{sec:one-dim-chiral-ferm}

\begin{figure}[b]
 \includegraphics[width=4.5cm]{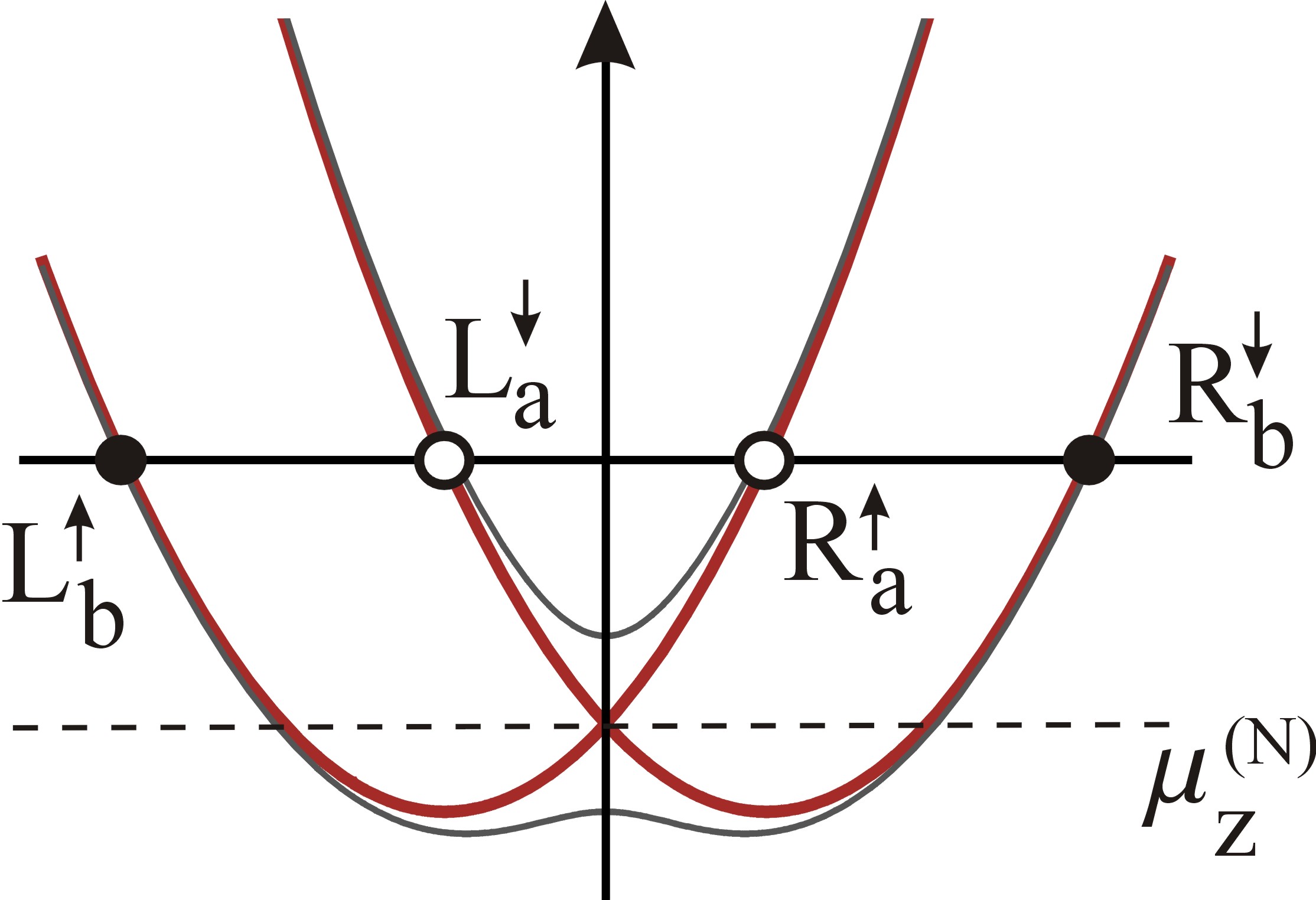}
 \caption{Chiral one-dimensional fermions in the spin-orbit quantum
   wire: $a$ and $b$ channels are shown by open and filled dots,
   resp. The magnetic field $B$ affects the dispersion relation for
   the $a$-channel only if the chemical potential is close to
   $\mu_z^{(N)}$.}
\label{fig:Bands}
\end{figure}

For strong spin-orbit coupling, $E_{\rm so} \gg B$, each band is
characterized by two Fermi momenta, 
\begin{equation}
k^{a/b}_{n} = \mp m\alpha + \sqrt{2m \left(\mu_z^{(N)}- \mu_z^{(n)} + E_{\rm so}\right)},
\end{equation}
as shown in Fig.~\ref{fig:Bands}. We aim to construct a low energy
Hamiltonian describing the system at energy scales $\lesssim E_{\rm so}$.  To this
end we represent the fields $\psi_\sigma^{(n)}$ as
\begin{eqnarray}
 \psi_{\uparrow}^{(n)} (x) &\simeq & R_{a\uparrow}^{(n)}(x)e^{ i k_{n}^a x} + 
L_{b\uparrow}^{(n)}(x)e^{-i k_{n}^b x}, \\
 \psi_{\downarrow}^{(n)} (x) &\simeq & L_{a\downarrow}^{(n)}(x)e^{ -i k_{n}^a x} + 
R_{b\downarrow}^{(n)}(x)e^{ i k_{n}^b x}, \label{eq:RL_channels}
\end{eqnarray}
in terms of a superposition of right ($R$) and left ($L$) chiral
fermions, and then linearize the Hamiltonian~(\ref{eq:H_approx_x})
around the Fermi momenta.  Modes with Fermi momenta of equal modulus
define a conduction channel.  We observe that each channel 
belongs to one of the two subsets, $a$ or $b$. In the $a$-channel
$R$-movers have spin up, and $L$-movers have spin down, while for the
channels of type $b$ spins are reversed.  The $a$-channel of band
index $N$ plays a distinguished role in that it has zero Fermi
momentum, $k_{N}^a=0$.  This particular mode defines what we call the
``topological channel". For ${\rm max}\{B,\Delta\} \ll E_{\rm so}$ it
is the only channel strongly susceptible to the magnetic field, thus
defining the topological phase of the whole wire. The effect of $B$
(but not $\Delta$) on other channels can be safely neglected.

We next collect all $R$ and $L$ fields into  two spinors,
\begin{eqnarray}
  \Psi &=& (R_{a\uparrow}, R_{b\downarrow}, L_{a\downarrow}, L_{b\uparrow}, 
  \bar R_{a\uparrow}, \bar R_{b\downarrow}, \bar L_{a\downarrow}, \bar L_{b\uparrow}),
  \label{eq:Psi:spinor}
\end{eqnarray}
and $\bar \Psi = (\sigma_x^{\rm ph}\Psi)^\mathrm{T}$ where the band index $(n)$
is left implicit.  The spinor $\Psi$ acts in an $8\times
N$-dimensional space, defined by the direct product 
of band (index $n$), channel ($a/b$), chiral ($R/L$) and
particle-hole spaces. In this representation the low energy Hamiltonian is given by
\begin{widetext}
\begin{equation}
{\cal H} = \sum_{n,m=1}^N\frac{1}{2}\int \bar\Psi^{(n)}(x)\left(
\begin{array}{cc}
\hat h_0^{(n)} \delta_{nm} + \hat W_{nm} & i \Delta\,\sigma_y^{RL}\otimes\sigma^{ab}_z \,\delta_{nm} \\
-i \Delta\, \sigma_y^{RL}\otimes\sigma^{ab}_z  \,\delta_{nm} & -(\hat h_0^{(n)} )^\mathrm{T}\delta_{nm} - \hat W^\mathrm{T}_{mn}
\end{array}\right)_{\rm ph}\Psi^{(m)}(x)\,\mathrm{d}x,
\label{eq:H_x}
\end{equation}
where 
\begin{equation}
\hat h_0^{(n)} = 
\left(
\begin{array}{cc}
-i v_n \partial_x - \mu & (B/2)\, ( \mathds{1}^{ab} + \sigma^{ab}_z)\, \delta_{Nn} \\
(B/2)\, ( \mathds{1}^{ab} + \sigma^{ab}_z) \, \delta_{Nn}  &  i v_n \partial_x - \mu
\end{array}\right)_{RL}.
\end{equation}
\end{widetext}
Here, $v_n = [2m (\mu_z^{(N)}- \mu_z^{(n)}+ E_{\rm so})]^{1/2}/m$ is
the Fermi velocity of the $n$-th channel, the chemical potential $\mu$
is now defined relative to the energy $\mu_z^{(N)}$, and $\hat
W_{nm}(x)$ is a random $8\times 8$ matrix satisfying the hermiticity
condition $(\hat W_{nm})^\dagger = \hat W_{mn}$.  In deriving this
Hamiltonian we have neglected oscillatory terms with phases $i x(\pm
k_a \pm k_b)$.  This is justified because the typical lengths involved
in the problem ($\xi \sim \Delta/v$ and $l_B \sim B/\alpha$) exceed by
far the Fermi length $\lambda_F \sim {\rm max}(1/k_a, 1/k_b)$
determined by the large spin-orbit energy $E_{\rm so}$.  We have also
used that the order parameter $\Delta(x)$ can be chosen to
be real.

\subsection{Majorana representation}
\label{sec:Majorana-rep}

The  
$8N\times 8N$ first-quantized matrix defining the Hamiltonian
\eqref{eq:H_x} satisfies to the p-h symmetry,
\begin{equation}
\hat H = -\sigma_x^{\rm ph} \hat H^\mathrm{T} \sigma_x^{\rm ph},
\end{equation}
which is the defining condition for a class $\mathrm{D}$ Hamiltonian
(here, the transposition acts on kinetic term as $\partial_x^\mathrm{T} = -\partial_x$.)
However, all what follows will be more conveniently formulated in an
alternative representation, in which the symmetry assumes a different
form: 
for each band $n$ we define a set of eight Majorana fields 
\begin{eqnarray}
\xi_{a}^{R} &=& (R_{a\uparrow} + \bar R_{a\uparrow})/\sqrt{2}, \quad 
\eta_{a}^{R} = (R_{a\uparrow} - \bar R_{a\uparrow})/\sqrt{2}i, \nonumber \\
\xi_{b}^{R} &=& (R_{b\downarrow} + \bar R_{b\downarrow})/\sqrt{2}, \quad 
\eta_{b}^{R} = (R_{b\downarrow} - \bar R_{b\downarrow})/\sqrt{2}i, \nonumber
\end{eqnarray}
(and analogous relations for the $L$--movers, with spin reversed), which we combine into 
the $8N$--spinor 
\begin{equation}
\tilde\chi = (\xi_{a}^{R}, \xi_{b}^{R}, \xi_{a}^{L},\xi_{b}^{L},
\eta_{a}^{R}, \eta_{b}^{R}, \eta_{a}^{L},\eta_{b}^{L})^\mathrm{T}. 
\end{equation}
The two spinors $\chi$ and $\Psi$ are related by the unitary transformation
\begin{eqnarray}
 \tilde\chi &=& U \Psi, \quad \tilde\chi^\mathrm{T} = \bar\Psi U^\dagger, \\
 U &=& \mathds{1}^{ab}\otimes \mathds{1}^{RL}\otimes \frac{1}{\sqrt{2}}\left(
\begin{array}{cc}
1 & 1 \\
i & -i
\end{array}\right)_{\rm ph}. \nonumber 
\end{eqnarray}
We combine the Majorana fields of $a$ and $b$ channels as  $\xi^{R} =
(\xi_{a}^{R},\xi_{b}^{R})^\mathrm{T}$, and reorder the spinor components as,
\begin{equation}
\chi = (\xi^R, \eta^L, \eta^R, \xi^L)^\mathrm{T}. \label{eq:chi_basis}
\end{equation}
In this representation (which will be used throughout the rest of the paper) 
the Hamiltonian~(\ref{eq:H_x}) takes the form
\begin{widetext}
\begin{align}
{\cal H} &= \sum_{n,m=1}^N\frac{1}{2}\int \chi^\mathrm{T}_{n}(x)\,\tilde{\mathcal{H}}_{nm}(x)\,\chi_{m}(x)\,\mathrm{d}x,\cr
&\tilde{\mathcal{H}}_{nm}(x)\equiv \left(
\begin{array}{cc}
\hat h_-^{(n)} \delta_{nm} +i  \hat W_{nm}^{--} & 
i \mu\,\sigma_z^{RL}\otimes\mathds{1}^{ab}\, \delta_{nm}  + i \hat W_{nm}^{-+}  \\
 - i \mu\,\sigma_z^{RL}\otimes\mathds{1}^{ab}\, \delta_{nm} + i \hat W_{nm}^{+-} & 
\hat h_+^{(n)} \delta_{nm} + i \hat W_{nm}^{++}
\end{array}\right),
\label{eq:H_Mb}
\end{align}
\end{widetext}
where the deterministic part reads
\begin{eqnarray}
\hat h^{(n)}_\pm &=& -i v_n\, \sigma_z^{RL} \otimes\mathds{1}^{ab}\,\partial_x - \sigma_y^{RL}\otimes\hat\Delta_\pm^{(n)},\\
\Delta_\pm^{(n)} &=& \Delta\,\sigma_z^{ab} \pm (B/2)( \mathds{1}^{ab} + \sigma^{ab}_z) \, \delta_{Nn},
\end{eqnarray}
and the random matrices are constructed as
\begin{eqnarray}
\hat W^{--} &=& \left(
\begin{array}{cc}
\hat w_2^{RR} & - \hat w_1^{RL} \\
\hat w_1^{LR} &  \hat w_2^{LL}
\end{array}
\right)_{ab}, \nonumber \\
\hat W^{-+} &=& \left(
\begin{array}{cc}
-\hat w_1^{RR} &  \hat w_2^{RL} \\
\hat w_2^{LR} &  \hat w_1^{LL}
\end{array}
\right)_{ab}, 
\end{eqnarray}
in terms of real (symmetric)
and imaginary (antisymmetric) parts of the random matrices  $(\hat W)^{RL} = \hat
w_1^{RL} + i\, \hat w_2^{RL}$ etc. The remaining blocks of the
$\hat W$-matrix are defined by  
\begin{equation}
\hat W^{+-} = -\sigma_z^{ab}\,\hat W^{-+}\,\sigma_z^{ab}, \qquad
\hat W^{++} =  \sigma_z^{ab}\,\hat W^{--}\,\sigma_z^{ab}.
\end{equation}
In the Majorana basis~(\ref{eq:chi_basis}), the class $\mathrm{D}$
symmetry is expressed through the antisymmetry ${\hat H}^\mathrm{T}=-{\hat H}$.

We finally specify the statistics of disorder. We choose $\hat W(x)$ to be a 
$\delta$-correlated and Gaussian distributed random  matrix of size $8N \times 8N$ with
a zero mean value $\langle \hat W(x) \rangle$ and variance 
\begin{align}
\!\!\!\langle w^{ij}_1(x) w^{i'j'}_1(x')\rangle =\frac{\gamma_w}{2}\delta(x-x')
(\delta_{ii'}\delta_{j'j'} + \delta_{ij'}\delta_{ji'}), \nonumber \\
\!\!\!\langle w^{ij}_2(x) w^{i'j'}_2(x')\rangle =\frac{\gamma_w}{2}\delta(x-x')
(\delta_{ii'}\delta_{jj'} - \delta_{ij'}\delta_{ji'}). \label{eq:g_w}
\end{align}
Here, the composite indices ($i,j$ etc.)  label states in the direct
product of band, channel and chiral spaces. The scattering matrices
defined in this way break time-reversal and spin rotation symmetry
(e.g., random spin-flip scattering caused by random spin-orbit terms
is included in (\ref{eq:g_w}).)  The strength of disorder set by the
coefficient $\gamma_w$ translates into the ``golden rule" scattering
rate
\begin{equation}
\tau^{-1} = 2\gamma_w^2 \sum_{n=1}^N (1/v_n)
\label{eq:golden_rule}
\end{equation}
of the normal conducting (i.e. superconductor decoupled) 
quantum wire. 

\section{Quasiclassical approach}
\label{quasi}

The kinetic term in the low energy Hamiltonian~(\ref{eq:H_Mb}) which
was discussed in the previous section is linear in momentum, and this
facilitates the formulation of quasiclassical equations of motion (aka
Eilenberger equations) for the model at hand~\cite{Eilenberger:1968}.
We here review the construction of these equations in a manner closely
following the spirit of
Refs.~\onlinecite{Shelankov:2000, Nazarov:1999}.  After introducing
the basics of the method (section~\ref{sec:eilenberger-method}) we
construct the Eilenberger $Q$--function in the limit of a single clean
``topological channel" (section~\ref{sec:PartIIIa}) and discuss the resulting 
density of states~(section~\ref{sec:clean-density-states}).  
In section~\ref{sec:mathdsz_2-topol-inva} we define the $\mathds{Z}_2$ topological invariant in terms of the
$Q$--matrix.  Section~\ref{sec:eilenb-equat-with} outlines the general construction of the
solution to the Eilenberger equation in the inhomogeneous disordered
wire with boundary conditions.

\subsection{Eilenberger method}
 \label{sec:eilenberger-method}

 We start by defining, 
\begin{equation}
g^{R/A}_{\epsilon,nm}(x,x') = \sqrt{v_n}\, G^{R/A}_{\epsilon,nm}(x,x')\, \sigma_z^{RL} \, \sqrt{v_m},
\label{eq:g_small}
\end{equation}
where $G_\epsilon^{R/A}(x,x') \equiv \langle x | (\epsilon \pm i0 - \tilde
 H)^{-1}| x' \rangle$ are the retarded and advanced Green's functions
 of the system. Under transposition (which in our current
 representation represents the particle-hole symmetry) the function
 $g$ behaves as 
\begin{equation}
g_\epsilon^R(x,x') = -\sigma_z^{RL}\left[ g_{-\epsilon}^A(x',x)\right]^\mathrm{T} \sigma_z^{RL},
\end{equation}
i.e. advanced and retarded Green's functions get interchanged. 
It is not hard to derive the two mutually adjoint differential equations 
\begin{eqnarray}
\partial_{x}\, g_\epsilon(x, x') +\mathcal{L}_\epsilon  g_\epsilon(x,
x')   &= -i \delta(x-x'),\cr
\label{eq:G1}
\partial_{x'} \, g_\epsilon(x, x') -   g_\epsilon(x, x')
\mathcal{L}_\epsilon  &= i\delta(x-x'),
\label{eq:6}
\end{eqnarray}
describing the dynamical evolution of $g$. Here, 
\begin{align}
  \label{eq:10}
  \mathcal{L}_\epsilon\equiv -i (\hat{\omega} - {\cal P} )
\end{align}
where matrix $\hat \omega$ has elements
\begin{equation}
(\hat{\omega})_{nm} \equiv (\epsilon/v_n)\, \sigma_z^{RL} \,\,\delta_{nm}
\end{equation} 
and the operator $\cal P$ 
is related to the Hamiltonian matrix $\tilde{\mathcal{H}}$ as
\begin{equation}
\mathcal{P}_{nm}= i\partial_x + v_n^{-1/2}\,(\sigma_3^{RL} \tilde H_{nm} )\, v_m^{-1/2}. \label{eq:L_def}
\end{equation} 
Due to the antisymmetry of $\tilde{\mathcal{H}}=-\tilde{\mathcal{H}}^\mathrm{T}$, the operator
$\mathcal{P}$ obeys the particle-hole symmetry
\begin{align}
  \label{eq:1}
  \mathcal{P}=-\sigma_z^{RL} \mathcal{P}^\mathrm{T} \sigma_z^{RL}.
\end{align}
We next define the Eilenberger function as
\begin{equation}
Q_\epsilon (x) = \lim_{x'\to x}\bigl[ 2i \,g_\epsilon(x,x') - {\rm sgn}(x-x')\bigr],
\label{eq:Q_def}
\end{equation}
where the subtraction of the $\mathrm{sgn}$-function regularizes a
discontinuity arising in $g$ at $x=x'$ due to the combination of
linear derivatives and $\delta$-function inhomogeneity in
Eqs.~\eqref{eq:G1}. Subtracting the two equations in \eqref{eq:G1}, we
then obtain the Eilenberger equation of motion
\begin{equation}
\partial_x Q_\epsilon(x) + \bigl[\mathcal{L}_\epsilon, Q_\epsilon(x)\bigr] = 0.
\label{Eq:Eil}
\end{equation}
The Eilenberger function $Q$ obeys the particle-hole symmetry
\begin{equation}
\sigma_z^{RL} \, Q_{-\epsilon}^\mathrm{T}(x)\, \sigma_z^{RL} = - Q_\epsilon(x),
\label{eq:Q_sym}
\end{equation}
and the normalization condition
$Q_\epsilon^2(x) = \mathds{1}$, where $\mathds{1}$ is the unit matrix (the
latter condition can be checked by verifying that Eq.~\eqref{Eq:Eil}
preserve the normalization $Q^2=\mathrm{const.}$). The unit-value of
the normalization constant is fixed by the jump height of the
$\mathrm{sgn}$-function in \eqref{eq:Q_def}.  We finally note that the
operator $\cal P$ can be straightforwardly  constructed from
\eqref{eq:L_def}, however, for our present purposes, we need not state
its explicit form in generality.  

\subsection{Eilenberger function in the clean limit}
\label{sec:PartIIIa}

As a warmup, we apply the quasiclassical approach to the limit ($\hat
W = 0$) of an infinite clean quantum wire subject to constant
$B,\Delta$. In this system all channels are decoupled, and we may
concentrate on the $4\times 4$ matrix $Q_\epsilon(B,\mu)$ describing
the ``topological" channel $a$ with $n=N$.  (The Eilenberger function
of the other channels may be obtained by setting $B=0$, rescaling the
velocity and transforming $\Delta \to -\Delta$ in the case of $b$-type
channels.)

We start by introducing the $4\times 4$
operator
\begin{equation}
L_\epsilon = -i 
\left(
\begin{array}{cc}
 \epsilon\, \sigma_z^{RL} - i\Delta_{-}\sigma_x^{RL} & -i\mu\, \mathds{1}^{RL} \\
  i\mu\,\mathds{1}^{RL} &  \epsilon\, \sigma_z^{RL} - i\Delta_{+}\sigma_x^{RL}
\end{array}
\right),
\label{eq:L_epsilon}
\end{equation}
as the reduction of the general operator $\mathcal{L}_\epsilon$ to a single channel.
The solution $Q_{\epsilon}$ is then determined by the relations
$[Q_{\epsilon},L_\epsilon]=0$ and $Q_{\epsilon}^2=\mathds{1}$. To solve
these equations, we assume $L_\epsilon$ to be diagonalized as 
\begin{align}
\label{eq:8}
L_\epsilon &= T (\hat \lambda\otimes \sigma_z^{RL})  T^{-1}, \nonumber \\ 
\hat \lambda &\equiv 
\left(
  \begin{matrix}
    \lambda_+&\cr&\lambda_-
  \end{matrix}
\right),
\end{align}
where the exact form of the (non-unitary) transformation matrix
$T$ will not be needed and 
\begin{eqnarray}
  \label{eq:4}
  \lambda_\pm &=&\sqrt{\lambda_0\pm \lambda},\nonumber \\
  \lambda_0 &=& B^2 + \Delta^2 - \mu^2 - \epsilon^2, \nonumber \\ 
  \lambda &=& 2\sqrt{ B^2\Delta^2 - (\Delta^2-\epsilon^2)\mu^2} 
\end{eqnarray}
are the eigenvalues. The defining equations for
$Q$ are then solved by matrices of the form 
\begin{equation}
Q_\epsilon = T\Lambda T^{-1}, 
\label{eq:Q_diag}
\end{equation}
where $\Lambda=(\pm 1,\dots,\pm 1)$ is a diagonal $4\times 4$ matrix containing
unit-modular entries in arbitrary configuration. The proper sign
structure 
is determined by causality, i.e. the sign of the infinitesimal offset $\epsilon\to
\epsilon\pm i0$ in the retarded/advanced Green function. That
increment enters in the combination $(\epsilon\pm i0)\sigma_z^{RL}$,
which means that the appropriate matrix structure of the retarded
Green's function (opposite for advanced) is given by
\begin{align}
  \label{eq:2}
  \quad \Lambda =\sigma_z^{RL}.
\end{align}
A more explicit derivation of this structure is detailed in Appendix
\ref{sec:eilenb-funct-q_eps}.

\begin{figure*}[t]
\begin{centering}
\includegraphics[width=16.0cm]{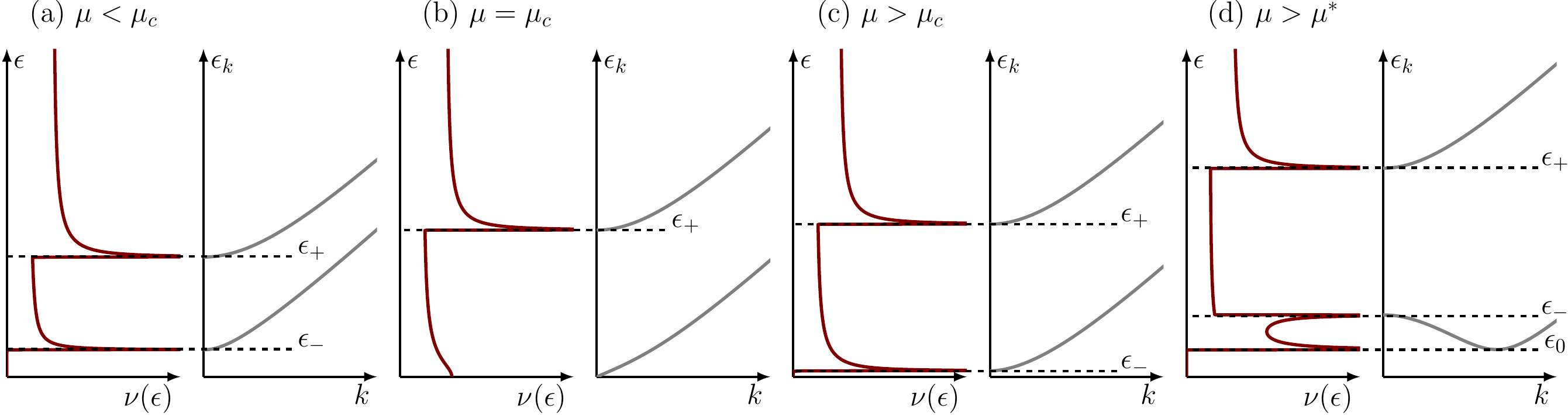}
\end{centering}
\caption{Sketch of the density of states and the corresponding
  dispersion relations in the clean nanowire. The gap is closed at the
  transition point $\mu=\mu_c$ (a), and opened again (b,c). For
  $\mu>\mu^*$ two minima develop (d).}
\label{fig:Dos_bulk}
\end{figure*}

\subsection{Clean density of states}
\label{sec:clean-density-states}

The density of states in the bulk of the topological wire is given by
$\nu(\epsilon)=(2\pi v)^{-1}\operatorname{Re}\operatorname{tr}
\sigma_z^{RL}Q_\epsilon$. The matrices $T$ diagonalizing $Q$ do not
commute with $\sigma_z^{RL}$, which means that a little extra work is
required to evaluate the trace. We start from the representation
\begin{align}
  Q_\epsilon = \frac{L_\epsilon}{\lambda_+}\, P^+ +  
\frac{L_\epsilon}{\lambda_-} \, P^-,
\label{eq:Q_mu}
\end{align}
where $P^+$ and $P^-$ are projectors on the space of $L_\epsilon$-eigenstates
with eigenvalues $\pm\lambda_+$ and $\pm\lambda_-$, resp.:
\begin{align}
  \label{eq:3}
  P^+&=T\mathrm{diag}(\mathds{1}_2,0)T^{-1},\cr
 P^-&=T\mathrm{diag}(0,\mathds{1}_2)T^{-1}.
\end{align}
That this representation faithfully represents the matrix
$Q_\epsilon$ is checked by application of \eqref{eq:Q_mu} in the
eigenbasis where all matrices assume a diagonal form. It remains to
obtain a representation of $P^\pm$ which does not make explicit
reference to the diagonalizing matrices $T$. To this end, notice that
(eigenrepresentation understood)
$L_\epsilon^2=\mathrm{diag}(\lambda_+^2 \mathds{1}_2,\lambda_-^2
\mathds{1}_2)=\lambda_0\mathds{1}_4+\lambda P^+-\lambda P^-$. This equation
can straightforwardly solved as
\begin{align}
  \label{eq:5}
  P^\pm={1\over 2}\left(\mathds{1}_4\pm {1\over \lambda} (L_\epsilon^2 - \lambda_0\mathds{1}_4)\right). 
\end{align}
Substituting this expression into \eqref{eq:Q_mu}, we obtain a
representation of $Q$ which makes reference only to the operator
\eqref{eq:L_epsilon}, and its eigenvalues. Computing the trace, we
obtain DoS profiles as shown in Fig.~\ref{fig:Dos_bulk}. Before
discussing the structure of these results, a general remark may be in
order: the DoS of a one dimensional quantum system is determined by an
interplay of the kinetic energy operator ($k\leftrightarrow
-i\partial_x$) and the ``potential" ($\mathcal{L}$). On the other hand,
we computed the DoS from $Q$ as determined
by $\mathcal{L}$, and this matrix seems to be oblivious to the 
kinetic energy. A closer look, however, shows that information on the band
dispersion sneaks in via the nonlinear constraint
$Q^2=\mathds{1}$. Indeed, the conservation of the constraint, and the unit
value of the normalization are consequences of the linearity of the
derivative operator in \eqref{eq:6}, which in this way co-determines
the structure of $Q$.

Inspection of Eq.~\eqref{eq:Q_mu} shows that the DoS contains
singularities at the zeros  $\lambda^\pm(\epsilon)=0$, which are
located at
\begin{align}
\epsilon_{\pm} = \vert B\pm \sqrt{\Delta^2 + \mu^2}\vert.
\end{align}
Let us assume that $B>\Delta$.
Fig.~\ref{fig:Dos_bulk}(a) shows the ensuing DoS profile, along with
the underlying dispersion relation for a value of the chemical
potential $\mu<\mu_c$, where 
\begin{align}
\mu_c = \sqrt{B^2 - \Delta^2},
\end{align}
defines a critical value where the lower of the DoS singularities,
$\epsilon_-$, touches zero and the band gap closes
[Fig.~\ref{fig:Dos_bulk}(b)]. At larger values $\mu>\mu_c$, the gap
reopens, (c), the DoS looks qualitatively similar to that of the
$\mu<\mu_c$ regime, but the system is in a topologically distinct
state (see the next section.) Finally, at values $\mu>\mu^\ast$, where
\begin{align}
\mu^* = \sqrt{B^2 + B \sqrt{B^2 + 4\Delta^2}}/\sqrt{2}, 
\end{align}
the lower band $\epsilon_-(k)$ develops an extremum at finite values
of $k$ which manifests in a third van-Hove singularity at the energy
\begin{align}
\epsilon_0 = \Delta\sqrt{1-B^2/\mu^2},
\end{align}
as shown in panel (d) of Fig.~\ref{fig:Dos_bulk}.

\subsection{$\mathds{Z}_2$ topological invariant}
\label{sec:mathdsz_2-topol-inva}

The symmetry Eq.~\eqref{eq:Q_sym} implies that at zero energy the product
$\sigma_z^{RL}Q_{\epsilon=0}$ is an antisymmetric $4\times 4$ matrix,
which implies the existence of a
Pfaffian. Due to the signature of $\Lambda$ the
determinant of $Q$ is unity, the same is true for the determinant of
the $4\times 4$ matrix 
$\sigma^{RL}_z=\sigma_z^{RL}\otimes \mathds{1}_2$. Consequently, the Pfaffian of
$\sigma_{z}^{RL}Q_{\epsilon=0}$ --- which squares to the determinant
of that matrix
--- may take one of two values, $\pm
1$. This motivates the definition of the topological index
\begin{align}
N_{\rm top} = {\rm Pf} (\sigma_z^{RL}\, Q_{\epsilon=0}) = \left\{
\begin{array}{ccc}
+1, &\quad& \mu > \mu_c \\
-1, &\quad& \mu < \mu_c
\end{array}
\right.,
\label{eq:Z2_inv}
\end{align} 
distinguishing between the two phases. Computing $N_{\mathrm{top}}$
from \eqref{eq:Q_mu}, we find the index structure stated in
\eqref{eq:Z2_inv}. (Right at the critical point $\mu=\mu_c$, the
matrix $Q_{\epsilon=0}$ becomes singular and the index cannot be
defined.)

\subsection{Eilenberger equation with disorder}
\label{sec:eilenb-equat-with}

In this section we discuss the formal solution of the Eilenberger
equation in the presence of disorder. The solution is ``formal" in the
sense that the Eilenberger Green function will be a functional of a
given realization of the disorder. To obtain practically useful
information, one will want to average over different realizations, and this
step of the computation needs to be done numerically, as discussed in
the next section. 

\begin{figure}[b]
\includegraphics[width=7.0cm]{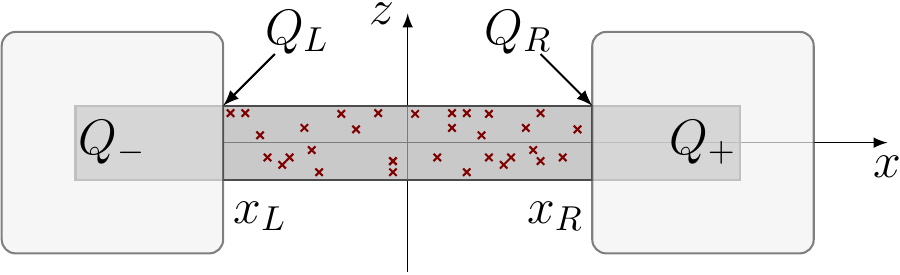}
\caption{Disordered spin-orbit wire connected to two ideal superconducting terminals which are described by the
Eilenberger functions $Q_+$ and $Q_-$. The $Q$-matrix at the boundaries between the  scattering
region and terminals is denoted by $Q_{R}$ and $Q_{L}$. In the superconductors
$Q$--matrix rapidly converges to either $Q_+$ or $Q_-$ on a scale of coherence 
length. }
\label{fig:Wire}
\end{figure}

To start with, consider the prototypical system
geometry shown in Fig.~\ref{fig:Wire}. The terminals indicated at the
left and right represent superconducting regions, assumed
non-disordered for simplicity. (This is an inconsequential assumption 
provided the rate of disorder scattering $\tau^{-1} \sim N \gamma_w^2\langle 1/v_n \rangle$ 
does not exceed the energy gaps ($\epsilon_-$ or $\epsilon_0$) in the
terminals.) In these regions, the Eilenberger
equation can be solved analytically, as discussed in the previous
section. Describing the disorder present in the center region in terms of a generalized
variant of the quasiclassical evolution operator $\mathcal{L}$, we
will show how the left and right asymptotic configuration of the Green
function get connected by a transfer matrix, $M$, functionally depending on
the disorder configuration. The ensuing generalized Green function
will then be the starting point for our numerical analysis.

To be more specific, we consider a quantum wire where the gap $\Delta(x)$ and/or chemical potential $\mu(x)$
vary in space in the region $|x|<L/2$ and saturate to some constants 
$\Delta_{L/R}$ and $\mu_{R/L}$ at $x\ll -L/2$ or $x\gg L/2$, respectively (Fig.~\ref{fig:Wire}).
These constants set asymptotic values of the $Q$-matrix,
\begin{equation}
Q_\epsilon(x\to -\infty) \equiv Q_{-}, \quad Q_\epsilon(x\to +\infty) \equiv Q_+,
\end{equation}
where $Q_{\pm}$ are constructed using the results of section~\ref{sec:PartIIIa}
for the homogeneous profile of $\Delta$, $B$ and $\mu$.
The boundary Green's function $Q_-$ and $Q_+$ may describe
different or equivalent topological phases of the wire.   

We denote the $Q$-matrices obtained at the interface between the
asymptotic superconducting regions, and the center disordered region,
resp., as 
\begin{equation}
 Q_R = Q(x_R), \quad Q_L=Q(x_L), 
\end{equation}
where $x_R=-x_L = L/2$.
These two configurations are related by a transfer matrix,
\begin{equation}
  Q_R = M(x_R, x_L)\, Q_L\, M^{-1}(x_R, x_L).
\label{eq:QL_QR_M}
\end{equation}
For arbitrary positions $x$ and $x'$ the formal expression for the
transfer matrix $M(x,x')$ at given energy $\epsilon$ follows from the
Eilenberger equation (\ref{Eq:Eil}),
\begin{equation}
M_\epsilon(x,x') = {\cal P}_x \exp\left\{i\int_{x}^{x'} \left[\hat\omega- {\cal P}(y)\right] \mathrm{d}y \right\},
\end{equation}
with ${\cal P}_x$ denoting the path-ordering operator. A relation
similar to Eq.~(\ref{eq:QL_QR_M}) connects the boundary matrices
$Q_{R/L}$ with the Green's function in the far right/left region of
the wire,
\begin{equation}
  Q_{\pm} = M(x, x_{R/L})\, Q_{R/L}\, M^{-1}(x,x_{R/L}),
\label{eq:g_Q_M}
\end{equation}
assuming that $x \to \pm \infty$.  The transfer matrix satisfies
certain symmetries. Along with the unitarity condition (cf. Eq.~\eqref{eq:L_def})
\begin{equation}
\label{eq:7}
\sigma_z^{RL} \,M_\epsilon^\dagger\,\sigma_z^{RL} = M_\epsilon^{-1},
\end{equation}
it also satisfies to the p-h symmetry:  with the use of
Eq.~\eqref{eq:L_def} one obtains
\begin{equation}
\mathcal{L}_\epsilon^\mathrm{T} = -\sigma_z^{RL} \mathcal{L}_{-\epsilon}\sigma_z^{RL},
\nonumber
\end{equation}
and this yields
\begin{equation}
\sigma_z^{RL}\, M_{-\epsilon}^\mathrm{T} \, \sigma_z^{RL} = M_{\epsilon}^{-1}.
\label{eq:SymM_D}
\end{equation}
We now aim to represent the $Q$-matrix in the scattering region in
terms of the transfer matrix $M_\epsilon$ and the asymptotic
Eilenberger functions $Q_\pm$.
We start by translating the transfer matrix relation \eqref{eq:g_Q_M}
to a set of algebraic conditions relating the matrices $Q_{R/L}$ to
$Q_\pm$. To this end, notice that the action of the
non-unitary (cf. Eq.~\eqref{eq:7}) transfer matrix on a generic
matrix $Q_R$ will in general produce exponentially increasing and
decreasing contributions. The former are inacceptable in that they lead to exponential divergencies
in the quasiclassical Green function. As is detailed in Appendix
\ref{sec:boundary-conditions}, the requirement of a non-divergent
Green function leads to the algebraic conditions
\begin{eqnarray}
&&(\mathds{1}+Q_-)(\mathds{1}-Q_L)=0,  \label{eq:Boundary_L1} \\
&&(\mathds{1}+Q_L)(\mathds{1}-Q_-)=0,  \nonumber
\end{eqnarray}
while the right matrix $Q_R$ should obey the analogous relations
\begin{eqnarray}
(\mathds{1}-Q_+)(\mathds{1}+Q_R)&=&0, \label{eq:Boundary_R1} \\
(\mathds{1}-Q_R)(\mathds{1}+Q_+)&=&0. \nonumber
\end{eqnarray}
We finally combined these equations with the transfer matrix relation
\eqref{eq:g_Q_M} to obtain closed expressions for $Q_{L/R}$ in terms
of the asymptotic configurations $Q_\pm$ and $M$. As a result
of a straightforward calculation detailed in Appendix
\ref{sec:boundary-conditions} we obtain
\begin{eqnarray}
Q_R &=& \mathds{1} + \frac{2}{Q_+ + M Q_- M^{-1}}\,(\mathds{1}- Q_+),  
\label{eq:Res_Q1}\\
Q_L &=& \mathds{1} + (\mathds{1}- Q_-)\,\frac{2}{Q_- + M^{-1} Q_+ M}, 
\label{eq:Res_Q2}
\end{eqnarray}
where $M\equiv M(x_R,x_L)$.

These formulae define the starting point for an efficient numerical
computation of the disordered Eilenberger function $Q(x)$.  To this
end, one computes the transfer matrix $M$ by numerical solution of
corresponding system of linear first order differential equations.
One next applies Eqs.~\eqref{eq:Res_Q1} and \eqref{eq:Res_Q2} to
obtain $Q_{R/L}$. Finally, our main object of interest, $Q(x)$, is
obtained by application of $M(x,x_{R/L})$ to either $Q_R$ or $Q_L$.

So far we have considered a quantum wire connected to two
superconducting terminals. However, the generalization of the method
to the system shown in Fig.~\ref{fig:Setup} is straightforward. The
key observation is that in the limit of vanishing barrier conductance
$g_T \ll 1$ the chiral fermion fields satisfy
\begin{equation}
R_{a\uparrow}(x_L) = e^{i\phi} L_{b\uparrow}(x_L),\quad R_{b\downarrow}(x_L) = e^{i\phi} L_{a\downarrow}(x_L),
\label{eq:reflection}
\end{equation}
where $\phi$ is (energy dependent) reflection phase shift. (Here, we
assumed the absence of barrier spin flip scattering, inter-channel
scattering or related complications.) In the limit of asymptotically
high potential barrier $\phi=\pi$.  Relations~(\ref{eq:reflection})
define boundary conditions for the Eilenberger function $Q_L$. As
verified in Appendix D, these conditions assume the form of
Eqs.~(\ref{eq:Boundary_L1}), where, however, the role of $Q_-$ is
taken by an ``effective" matrix $Q_-\equiv Q_-(\phi)$ describing the
tunnel junction.  The explicit form of this matrix reads
\begin{equation}
  Q_- = (\sin(\phi)\sigma_z^{\rm ph} + 
\cos(\phi)\sigma_x^{\rm ph} )\otimes \sigma_x^{RL}\otimes\sigma_x^{ab}.
\label{eq:g_Tjunction}
\end{equation}
This matrix also satisfies $Q_-^2=\mathds{1}$, and the
relations~(\ref{eq:Res_Q1},\ref{eq:Res_Q2}) stay intact.

We close this section with an important statement: the Eilenberger
functions $Q_+$ and $Q_-$ of the terminals define the ``Majorana
number"~\cite{Kitaev:2001} of the wire as
\begin{equation}
{\cal M} = {\rm Pf} (\sigma_z^{RL}\, Q_+) \, {\rm Pf} (\sigma_z^{RL}\, Q_-), 
\end{equation}
in terms of the product of two $\mathds{Z}_2$ topological invariants
given by Eq.~(\ref{eq:Z2_inv}).  It is shown in Appendix
\ref{sec:majorana-mode}, that for ${\cal M}=-1$ the system supports a
Majorana fermion localized in the scattering region between two
terminals.

\section{Numerics}
\label{num}

We next turn to the discussion of our numerical results obtained for
the setup shown in Fig.~\ref{fig:Setup}. We have solved
the quasiclassical equations according to the algorithm of
section~\ref{sec:eilenb-equat-with} and from there computed the local
density of states (LDoS) 
$\nu_L(\epsilon) = (2\pi v)^{-1} \operatorname{Re}\, {\rm tr}(\sigma_z^{RL} Q_L)$
at the left end of the wire close to the tunnel barrier.  In the
actual calculations we shifted the energy into the complex plane,
$\epsilon \to \epsilon + i \Gamma$.  This shift accounts for the fact
that in the ``real'' system states may escape to a continuum of lead
scattering states, which gives them the status of quantum resonances
of finite lifetime $\sim \Gamma^{-1}$. The corresponding decay
rate~\cite{Wimmer:2011} is given by the standard golden rule
expression, $\Gamma \sim g_T \delta$ where $\delta \sim \pi v/ 2 N L$
is the mean level spacing in the scattering region of size $L$. In principle,
one might numerically \textit{compute} the broadening by an extension
of the numerical setup so as to include and extended normal metallic
scattering region to the left of the tunnel barrier. This, however,
would slow the performance of the numerics which is why we prefer to
introduce the broadening ``by hand". At any rate, the Majorana peak,
present for ${\cal M}=-1$, will acquire a finite width $\sim
\Gamma$. The DoS profiles obtained as a result of the procedure
outlined above are presented and discussed in section
\ref{sec:qualt-disc-results} above.

\section{Conclusions}
\label{concl}

In this paper we have adapted the Eilenberger quasiclassical approach to the specific conditions of a class $\mathrm{D}$
topological quantum wire. The most striking feature of the
quasiclassical approach is that the Green
functions of the system are described by ordinary, and hence easily
solvable differential equations. A numerical solution of these
equations for given realizations of the disorder produces 
accurate information on the spectral properties of reasonably
realistic model systems hosting Majorana boundary states. Our results confirm
the predictions made in Ref. \cite{Bagrets:2012} for a more idealized
system, viz. that disorder generates spectral peaks which can be
confused for genuine Majorana peaks. It thus seems that an unambiguous
detection of the Majorana might call for a measurement scheme beyond
direct tunneling spectroscopy.

\section{Acknowledgments}
\label{sec:acknowledgments}

We thank C.~W.~J.~Beenakker, F.~von Oppen and M.~Feigel'man for fruitful discussions. 
This work was supported by the collaborative research grant  SFB/TR12 of the Deutsche Forschungsgemeinschaft.

\appendix

\section{Derivation of Eq.~\eqref{eq:2}}
\label{sec:eilenb-funct-q_eps}

The goal of this Appendix is to derive the structure
\eqref{eq:2} by explicit calculation. Our starting point is the
momentum representation of the equations of motion for the
quasiclassical Green function $g(x,x';\epsilon)\equiv g(x-x';\epsilon)$,
\begin{equation}
\left( - p + \frac{i}{v} L_\epsilon \right)\,g(p,\epsilon)=1.
\end{equation}
Assuming a diagonal representation as in \eqref{eq:8}, we obtain
\begin{equation}
g(p,\epsilon) = - \,T {1\over p-i{\hat \lambda\over v}\otimes \sigma_z^{RL}} T^{-1}.
\label{eq:Q_eigen}
\end{equation}
The inspection of eigenvalues $\lambda_\pm$ shows that for the retarded Green's function, i.e.
at $\epsilon\to \epsilon + i 0$, one has ${\rm Re}(\lambda_\pm)>0$. This gives
\begin{align}
\label{eq:9}
g(x,x';\epsilon)&=T\int {dp\over 2\pi}\frac{e^{ip x}}{p - i {\hat
    \lambda\over v} \otimes
  \sigma_z^{RL}} T^{-1}= \\
&=-{i\over 2} T (\sigma_z^{RL} +\mathrm{sgn} (x-x')) \,e^{-{\hat
  \lambda \over v}|x-x'|}\, T^{-1}.\nonumber
\end{align}
Sending $x'\to x$ and comparing to \eqref{eq:Q_def}, we obtain the
identifications \eqref{eq:Q_diag} and \eqref{eq:2}. 
 
\section{Boundary conditions}
\label{sec:boundary-conditions}

In this Appendix we derive the boundary
conditions~(\ref{eq:Boundary_L1},\ref{eq:Boundary_R1}) for the
$Q$-matrix.  The method we use is adapted from the circuit theory of
Refs.~\cite{Nazarov:1999,Cottet:2009}. For simplicity, we consider the
$4\times 4$ matrices describing individual channels, generalization to
many coupled channels is straightforward.

Consider the quasiclassical
function $g(x,x')\equiv g(x,x';\epsilon)$ to the right of the right terminal, 
$x,x'\ge x_R$. By assumption, the generator
$L_\epsilon\big|_{x>x_R}$ is constant in space, and this means that the
equations \eqref{eq:6} defining $g$ admit the solutions 
\begin{align*}
g(x,x_R) &= e^{-L_\epsilon (x-x_R)} g(x_R+0,x_R),\cr
g(x_R,x) &= g(x_R,x_R+0)e^{L_\epsilon (x-x_R)},
\end{align*}
where we have again set $v=1$ for notational simplicity.
According to Eq.~(\ref{eq:9}), we have  
\begin{align}
g(x_R+0,x_R) &= -{i\over 2}(Q_R + \mathds{1}),\nonumber \\
\label{eq:g_Qpm}
g(x_R,x_R+0) &= -{i\over 2}(Q_R - \mathds{1}),
\end{align}
so that
\begin{align*}
g(x,x_R) &= -{i\over 2}e^{-L_\epsilon (x-x_R)} (Q_R + \mathds{1}),\cr
g(x_R,x) &= -{i\over 2} (Q_R + \mathds{1})  e^{L_\epsilon (x-x_R)}.
\end{align*}
Defining $\bar g = T^{-1} g T$ and $\bar Q_R = T^{-1} Q_R T$, we next
transform to the representation \eqref{eq:8} to arrive at
\begin{align*}
\bar g(x,x_R) &= -{i\over 2}e^{-\hat \lambda \otimes \sigma_3^{RL}
  (x-x_R)} (\bar Q_R + \mathds{1}),\cr
\bar g(x_R,x) &= -{i\over 2} (\bar Q_R + \mathds{1})  e^{\hat \lambda \otimes
  \sigma_3^{RL} (x-x_R)}.
\end{align*}
The key observation now is that the matrix functions $\bar g$ must
remain finite as $x\to \infty$. Due do the positivity of the matrix
$\hat \lambda$, 
this is equivalent to the condition $(\mathds{1}-\sigma_3^{RL})(\hat
Q_R+\mathds{1}) = (\mathds{1}-\sigma_3^{RL}\otimes \mathds{1})(\hat
Q_R+\mathds{1}) =(\mathds{1}-\Lambda)(\hat
Q_R+\mathds{1}) =0$. By the same token we have $(\bar Q_R +
\mathds{1})(\mathds{1}+\Lambda)=0$. Finally, noting that $T\Lambda
T^{-1}=Q_+$ represents the asymptotic form of the $Q$-matrix deep
in the superconductor, we arrive at
Eqs.~\eqref{eq:Boundary_R1}. Eqs.~\eqref{eq:Boundary_L1} are shown in
an analogous way. 

We are now in position to derive Eqs.~\eqref{eq:Res_Q1} and
\eqref{eq:Res_Q2}. To this end, we take Eqs.~(\ref{eq:Boundary_L1})
and multiply it by the transfer matrix $M$ from the left and by $M^{-1}$ from the right. Bearing in mind Eq.~(\ref{eq:g_Q_M}),
one obtains
\begin{equation}
\mathds{1} - M Q_- M^{-1}\, Q_R + M Q_- M^{-1} - Q_+ = 0.
\end{equation}
Adding this relation to Eqs.~(\ref{eq:Boundary_R1}), we arrive at
\begin{equation}
2\cdot\mathds{1}-M Q_- M^{-1}\, Q_R - Q_+ Q_R + M Q_- M^{-1} - Q_+ = 0,
\end{equation}
which in one more step yields the
result~(\ref{eq:Res_Q1}). Eq.~\eqref{eq:Res_Q2} is proven
analogously.

\section{Majorana mode}
\label{sec:majorana-mode}

In this Appendix we discuss the ``Majorana number" ${\cal M}$ and show
how the Majorana state emerges from the quasiclassical Eilenberger
function.

We consider the spectrum $\{E_j\}$ of Andreev bound states which follows from the poles of the 
$Q$-function~(\ref{eq:Res_Q1},\ref{eq:Res_Q2}).
If we denote 
\begin{equation}
{\cal D}(\epsilon) = Q_+(\epsilon) + M(\epsilon) Q_-(\epsilon) M^{-1}(\epsilon),
\label{eq:D_matrix}
\end{equation}
then the energies $E_j$ are solutions of the secular equation ${\rm det}\,{\cal D}(E_j)=0$.
The Majorana state, if it exists, corresponds to $E_0=0$.

To proceed, we introduce matrices
$\tilde Q_{\pm}(\epsilon) = Q_{\pm}(\epsilon)\sigma_z^{\rm RL}$,
and the secular matrix
$\tilde {\cal D}(\epsilon) = {\cal D}(\epsilon) \sigma_z^{\rm RL}$. 
Since in the subgap interval of energies there is no distinction
between the retarded and advanced Green's function, the unitarity relation
$(G^{R/A})^\dagger
=G^{A/R}$, the basic definitions Eqs.~\eqref{eq:g_small} and
\eqref{eq:Q_def} imply 
$\tilde Q_{\pm}^\dagger(\epsilon) = -\tilde Q_{\pm}(\epsilon)$.
Further, particle-hole symmetry (\ref{eq:Q_sym}) yields
$\tilde Q_{\pm }^\mathrm{T}(-\epsilon) = -\tilde Q_\pm(\epsilon)$.
Taking into account the class $\mathrm{D}$ symmetry of the transfer
matrix, Eq.~(\ref{eq:SymM_D}), the secular matrix takes a form
\begin{equation}
\tilde {\cal D}(\epsilon) = \tilde  Q_+(\epsilon) + M(\epsilon) \tilde Q_-(\epsilon) M^\mathrm{T}(-\epsilon).
\end{equation} 
It satisfies 
$\tilde {\cal D}^\mathrm{T}(\epsilon) = - \tilde {\cal D}(-\epsilon)$,
and thereby guarantees that Andreev bound states appear in 
pairs $\pm E_j$.

One may ask now whether a zero energy solution of the secular equation exists or not. At
$\epsilon=0$, the particle hole symmetry puts tighter restrictions on
the matrices $\tilde Q_{\pm}$, $M$ and $\tilde {\cal D}$ (we omit
the energy argument for brevity). One gets
\begin{eqnarray}
&&\tilde Q_{\pm}^\mathrm{T} = - \tilde Q_{\pm}, \quad \tilde Q_{\pm}^* = \tilde Q_{\pm}, \\
&&\sigma_z^{\rm RL} M^\mathrm{T}\sigma_z^{\rm RL} = M^{-1}, \quad M^* = M, \label{eq:Sym_E0}\\
&&\tilde {\cal D}^\mathrm{T} = - \tilde{\cal D}, \quad \tilde {\cal D}^* = \tilde{\cal D}. 
\end{eqnarray} 
We thus observe that both $\tilde Q_{\pm}$ and $\tilde {\cal D}$ are real
antisymmetric matrices of size $8N\times 8N$, which enables us to
rewrite the secular equation in terms of a Pfaffian
\begin{equation}
{\rm Det} \tilde {\cal D} = \left[ {\rm Pf}(\tilde {\cal D})\right]^2 = 
\left[ {\rm Pf}(\tilde Q_+ + M \tilde Q_- M^\mathrm{T})\right]^2 = 0.
\end{equation}
Let us denote by $\Omega^L_0$ the left null space of the matrix
$\tilde {\cal D}$, i.e. any bra $\langle \phi| \in \Omega^L_0$ by
definition satisfies $\langle \phi| \tilde {\cal D} = 0$.  Since $\tilde
{\cal D}$ is antisymmetric, the dimension of its null space is even, ${\rm
  dim}\, \Omega^L_0 = 2 {\cal N}$.  
At this stage we make use of a mathematical lemma proven  in 
Appendix A of Ref.~\onlinecite{Fulga:2011}:  the parity of the number ${\cal N}$
can be expressed as
\begin{equation}
(-1)^{\cal N} = {\rm Pf}\,\tilde Q_+\, {\rm Pf}\,[M \, \tilde Q_-\, M^\mathrm{T}] =   
{\rm Det}\, M\, {\rm Pf}\,\tilde Q_-\, {\rm Pf}\,\tilde Q_+. 
\end{equation}
The particle-hole symmetry~(\ref{eq:Sym_E0}) implies that ${\rm Det}\,
M = \pm 1$ at zero energy.  Now, the transfer matrix $M(x,x')$ is a
continuous function of its arguments, with initial value $M(x=x') =
\mathds{1}$. We thus conclude that ${\rm Det}\, M = + 1$, so that the
parity of ${\cal N}$ is determined by the terminal configurations,
\begin{equation}
(-1)^{\cal N} = {\rm Pf}\,\tilde Q_-\, {\rm Pf}\, \tilde Q_+.
\end{equation}
This parity is equal to the ``Majorana number" ${\cal M}$ introduced in section~\ref{sec:eilenb-equat-with}.

Let us now focus on the most interesting case ${\cal N}=1$,
corresponding, as we will see, to a single Majorana
mode~\footnote{In case of odd ${\cal
    N}\ge 3$ as well as non-zero even ${\cal N}\ge 2$, the ${\cal
    N}$-fold degeneracy of a zero level is accidental and not
  topologically protected. It can be reduced down to ${\cal N}=1$ or
  $0$ by continuous distortion of the transfer matrix $M$}. In this
case we have ${\rm dim}\, \Omega^L_0 =2$, and the null space is
spanned by two linearly independent vectors. Let $\langle \phi_1| \in
\Omega^L_0$ be the first basis vector. It is easy to check that $\langle
\phi_2| = \langle \phi_1| Q_+ \in \Omega^L_0$, and we may choose this
state for the second basis vector in $\Omega_0^L$. Indeed, for any
$\langle \phi_1| \in \Omega^L_0$ one has $\langle \phi_1| \tilde {\cal
  D} = \langle \phi_1| {\cal D} = 0$. Using the definition of ${\cal
  D}$, Eq.~(\ref{eq:D_matrix}), we deduce that
\begin{equation}
\langle \phi_1| Q_+ M = - \langle \phi_1| M Q_-.
\end{equation}
This relation enables us to evaluate $\langle \phi_2| {\cal D}$ as
\begin{eqnarray}
\langle \phi_2| {\cal D} &=&  \langle \phi_1| \left( \mathds{1} + Q_+ M Q_- M^{-1}\right) \nonumber \\
&=& \langle \phi_1|  + \left(  \langle \phi_1|  Q_+ M \right)\, Q_- M^{-1} \nonumber \\
&=&  \langle \phi_1|- \langle \phi_1|  M Q_- Q_- M^{-1} = 0,
\end{eqnarray}
and hence we proved that $\langle \phi_2| \in \Omega_0^L$. Since
${\cal D}$ is a real antisymmetric matrix, its right null space is
obtained as $\Omega_0^R = (\Omega_0^L)^\dagger$.  In other words, the
two kets $|\phi_{1,2}\rangle = (\langle \phi_{1,2})|)^\dagger$ satisfy
the relation $\tilde {\cal D} |\phi_{1,2}\rangle = 0$.

Let us look at the Eilenberger function $Q_R(\epsilon)$ around its
pole at $\epsilon=0$. According to Eq.~\eqref{eq:Res_Q1}, it
can be represented by two equivalent equations,
\begin{eqnarray}
Q_R &=& \mathds{1} + 2{\cal D}^{-1} (\mathds{1} -Q_+), \\
Q_R &=& -\mathds{1} + 2(\mathds{1} + Q_+){\cal D}^{-1}.
\end{eqnarray} 
Multiplication by $\sigma_3^{\rm RL}$ yields 
\begin{eqnarray}
\sigma_3^{\rm RL} \tilde Q_R \sigma_3^{\rm RL} &=& \sigma_3^{\rm RL}+ 2\tilde {\cal D}^{-1} (\mathds{1} -Q_+), \\
\sigma_3^{\rm RL} \tilde Q_R \sigma_3^{\rm RL}  &=& -\sigma_3^{\rm RL}+ 2(\mathds{1} - Q_+^\mathrm{T})\tilde{\cal D}^{-1}.
\end{eqnarray} 
At $\epsilon \to 0$ the inverse operator from the secular matrix has
the pole structure, $\tilde{{\cal D}}^{-1} \sim \mathcal{R}/\epsilon$, where the
matrix $\mathcal{R}$ is its residue at zero energy.

At this stage it is advantageous to introduce a new bra 
\begin{equation}
\langle \chi_\pm| = \langle \phi_1 |\pm \langle \phi_2|=\langle \phi_1|(\mathds{1} \pm Q_+),
\end{equation}
 and ket
\begin{equation}
|\chi_\pm \rangle  = (\mathds{1} \pm Q_+^\mathrm{T}) |\phi_1\rangle,
\end{equation}
basis in the null spaces $\Omega_0^L$ and $\Omega_0^R$, resp. The bra
basis $\langle \chi_\pm|$ is not orthogonal and we can denote by
$|\eta_\pm\rangle$ the ket basis, which is dual to it, $\langle
\chi_\sigma| \eta_{\sigma'}\rangle = \delta_{\sigma\sigma'}$.
Using these definitions, we may formulate resolutions of unity,
\begin{align*}
  \mathds{1}=|\eta_\sigma\rangle \langle \chi_\sigma| =
  |\chi_\sigma\rangle \langle \eta_\sigma|,
\end{align*}
where $\sigma=\pm$ is summed over. 

Now let us look at the singular part of the
matrix $\tilde Q_R$, 
\begin{eqnarray}
\sigma_z^{RL} \tilde Q_R^{\rm sing} \sigma_z^{RL} &=&
\frac{2}{\epsilon}\mathcal{R}  (\mathds{1} - Q_+), \nonumber \\
&=& \frac{2}{\epsilon} (\mathds{1} - Q_+^\mathrm{T})\mathcal{R}. 
\label{eq:Q_sing}
\end{eqnarray}
We note that the matrix $P_-={1\over 2}(\mathds{1} - Q_+)$ acts as the projector in the space $\Omega_0^L$, since
\begin{equation}
\langle \chi_+| P_- = 0,\quad \mbox{and} \quad \langle \chi_-| P_-~=~\langle \chi_-|.
\end{equation}
Similar relations hold in the ket space $\Omega_0^R$:
\begin{equation}
P_-^\mathrm{T} |\chi_+\rangle =0, \quad \mbox{and} \quad P_-^\mathrm{T} |\chi_-\rangle =  |\chi_-\rangle. 
\end{equation}
Equivalently, we may write 
\begin{align*}
  P_- = |\eta_-\rangle\langle \chi_-|,\qquad P_-^\mathrm{T} = |\chi_-\rangle\langle\eta_-|.
\end{align*}
Using these properties, it follows $\sigma_z^{RL} \tilde Q_R^{\rm
  sing} \sigma_z^{RL}=\tfrac{4}{\epsilon} \mathcal{R} P_-=\tfrac{4}{\epsilon}
P_-^\mathrm{T} \mathcal{R}$. Multiplying  these relations by $P_-$ from
the right, and using the projector property $P_-^2=P_-$, we obtain the
relation
\begin{align*}
  \sigma_z^{RL} \tilde Q_R^{\rm
  sing} \sigma_z^{RL}={4\over \epsilon} P_-^\mathrm{T} \mathcal{R} P_-={4\over
  \epsilon}  |\chi_-\rangle\langle\eta_-|\mathcal{R} |\eta_-\rangle\langle \chi_-|,
\end{align*}
or
\begin{align*}
   \tilde Q_R^{\rm
  sing} ={4 \mathcal{R}_{--} \over \epsilon} \sigma_z^{RL}  |\chi_-\rangle
\langle \chi_-|\sigma_z^{RL},
\end{align*}
where we defined $\mathcal{R}_{--}\equiv \langle\eta_-|\mathcal{R}
|\eta_-\rangle$.  Taking into account our definition of the
quasiclassical Green's function, Eqs.~(\ref{eq:g_small}) and
(\ref{eq:Q_def}), we finally obtain that the singular part of the
propagator around zero energy at $x=x'=x_R$ takes the form
\begin{equation}
G(x_R,x_R;\epsilon)  \sim   \left(\frac {2 \mathcal{R}_{--}}{i\epsilon}\right)
\hat v^{-1/2}\sigma_z^{RL}| \chi_- \rangle
\langle \chi_-| \sigma_z^{RL} \hat v^{-1/2},
\end{equation}
where $\hat v$ is the diagonal velocity matrix.

This expression can be compared with the spectral decomposition of the Green's function,
\begin{align*}
G(x,x',\epsilon) &=&{|\psi_0(x)\rangle \langle \psi_0(x')|\over \epsilon}+\sum_{j}
\frac{|\psi_{E_j}(x)\rangle\langle \psi_{E_j}(x')|}{\epsilon-E_j} \\
&+& \left( \int_{E_g}^{+\infty} \mathrm{d}E + \int_{-\infty}^{-E_g}\mathrm{d}E \right) 
\frac{|\psi_{E}(x)\rangle\langle \psi_{E}(x')|}{\epsilon-E},\nonumber \\
\nonumber
\end{align*}
where $|\psi_{E}(x)\rangle$ are normalized
eigenfunctions of the BdG Hamiltonian, $E_g$ is the gap in the spectrum of the wire, the sum is going
over the set of subgap Andreev levels (we have particle-hole symmetry
$E_{-j} = - E_j$) and the first term is present if the system
contains a Majorana state.  One thus concludes that
\begin{equation}
\left(\frac{2\mathcal{R}^{--}}{i\hat v}\right)^{-1/2}\sigma_z^{RL}| \chi_- \rangle = |\psi_{0}(x_R)\rangle
\end{equation}
is the amplitude of the Majorana particle at the point $x_R$. 
The amplitude of the Majorana state at any other point $x$ can be obtained
from $|\psi_{0}(x_R)\rangle$ by applying the transfer matrix $M(x,x_R)$.

To conclude, we have shown that if the Majorana number of our system is topologically non-trivial, 
i.e. ${\cal M}=-1$, then the Green's function has the pole at zero energy. Up to a prefactor, 
the residue at zero energy is then the projector on the one dimensional linear subspace spanned 
by the Majorana particle.

\section{Matrix $Q_-$ of a tunnel junction}
\label{sec:tunnel_junction}

In this appendix we show that the boundary conditions for the
Eilenberger matrix $Q$ of a spin-orbit wire terminated at the left end
by a (infinitely high) tunnel barrier are equivalent to algebraic
relations~(\ref{eq:Boundary_L1}) with the effective matrix $Q_-$ given
by Eq.~(\ref{eq:g_Tjunction}).

Let us work in the original particle-hole basis where the spinor
$\Psi$ takes the form specified by Eq.~(\ref{eq:Psi:spinor}). We start
by rewriting the left boundary conditions~(\ref{eq:reflection}) in a
matrix form. If one defines the reflection matrix
\begin{equation}
\hat r = \left(
\begin{array}{cc}
0 & e^{i\phi} \sigma_x^{RL} \\
e^{-i\phi} \sigma_x^{RL} & 0
\end{array}
\right)
\end{equation}
in the particle space, then $\hat r^*$ is the reflection matrix for holes.
Consequently, the spinor $\Psi(x_L)$ satisfies the condition
\begin{equation}
\Bigl(\mathds{1}-\hat R\Bigr) \Psi(x_L) = 0, \qquad 
\hat R = \left(
\begin{array}{cc}
\hat r & 0 \\
0 & \hat r^*
\end{array}
\right).
\end{equation}
The full reflection matrix $\hat R$ is unitary obeying the
particle-hole  symmetry,
$\sigma_x^{\rm ph} R^\mathrm{T} \sigma_x^{\rm ph} = R$. Hence the boundary condition for
the bar spinor $\bar \Psi = (\sigma_x^{\rm ph} \Psi)^\mathrm{T}$ takes the similar form
\begin{equation}
\bar \Psi(x_L) \Bigl(\mathds{1}-\hat R\Bigr)  = 0. 
\end{equation}
According to the definition~(\ref{eq:g_small}), the Green's function $g_e(x,x')$ inherits  
the boundary condition right to the tunnel junction from those of the direct product of two
spinors, $\Psi(x) \otimes \left(\bar{\Psi}(x')\, \sigma_z^{RL}\right)$. We thus conclude that
\begin{align*}
\bigl(\mathds{1}- \hat R\bigr) g(x_L,x;\epsilon) &= 0, \\
g(x,x_L;\epsilon) \bigl(\mathds{1} + \hat R\bigr) &= 0. 
\end{align*}
When deriving the second relation, we have taken into account that 
$\sigma_z^{RL}\, \hat r\, \sigma_z^{RL} = -\hat r$. The above conditions are valid for any $x>x_L$.
Sending now $x \to x_L + 0$ and using the relations~(\ref{eq:g_Qpm}) written for the matrix $Q_L$, 
one obtains
\begin{eqnarray*}
&&(\mathds{1}- \hat R)(\mathds{1}-Q_L)=0, \\
&&(\mathds{1}+Q_L)(\mathds{1}-\hat R)=0. 
\end{eqnarray*}
We see that these boundary conditions are equivalent to the algebraic conditions~(\ref{eq:Boundary_L1})
if one identifies $Q_- = -\hat R$. The normalization $\hat R^2 = \mathds{1}$ guaranties that
$Q_-$ belongs to the manifold of the quasiclassical Eilenberger functions. Transforming matrix $Q_-$
into the Majorana representation~(\ref{eq:chi_basis}) we finally obtain the result~(\ref{eq:g_Tjunction}).

The scattering phase $\phi$ is the parameter of the matrix $Q_-$ which may depend on the 
band index $n$ and characterizes the tunnel junction. In our numerical simulations we have used
$\phi=\pi$ for all bands, which corresponds to the infinitely high barrier as compared to the
energies of Andreev bound states.


\bibliography{my_biblio}
\bibliographystyle{apsrev4-1}

\end{document}